\documentclass[twocolumn,twocolappendix]{aastex701}

\usepackage{comment}
\shorttitle{Deep Rest-UV Grating Spectra of LRDs}
\shortauthors{Tang et al.}

\begin{document}

\title{SPURS: Evidence for Clumpy Neutral Envelopes and Ionized IGM Surrounding Little Red Dots in Abell 2744 from Ultra-Deep Rest-UV Spectroscopy}

\author[0000-0001-5940-338X]{Mengtao Tang}
\affiliation{Tsung-Dao Lee Institute, Shanghai Jiao Tong University, 1 Lisuo Road, Shanghai 201210, People’s Republic of China}
\affiliation{School of Physics and Astronomy, Shanghai Jiao Tong University, 800 Dongchuan Road, Shanghai 200240, People’s Republic of China}
\affiliation{State Key Laboratory of Dark Matter Physics, Shanghai Jiao Tong University, 1 Lisuo Road, Shanghai 201210, People’s Republic of China}
\email[show]{mengtao.tang@sjtu.edu.cn}

\author[0000-0001-6106-5172]{Daniel P. Stark}
\affiliation{Department of Astronomy, University of California, Berkeley, Berkeley, CA 94720, USA}
\email{dpstark@berkeley.edu}

\author[0000-0002-3407-1785]{Charlotte A. Mason}
\affiliation{Cosmic Dawn Center (DAWN)}
\affiliation{Niels Bohr Institute, University of Copenhagen, Jagtvej 128, 2200 Copenhagen N, Denmark}
\email{charlotte.mason@nbi.ku.dk}

\author[0000-0002-2178-5471]{Zuyi Chen}
\affiliation{Cosmic Dawn Center (DAWN)}
\affiliation{Niels Bohr Institute, University of Copenhagen, Jagtvej 128, 2200 Copenhagen N, Denmark}
\email{zuyi.chen@nbi.ku.dk}

\author[0000-0003-1561-3814]{Harley Katz}
\affiliation{Department of Astronomy \& Astrophysics, University of Chicago, 5640 S Ellis Avenue, Chicago, IL 60637, USA}
\affiliation{Kavli Institute for Cosmological Physics, University of Chicago, Chicago IL 60637, USA}
\email{harleykatz@uchicago.edu}

\author[0000-0003-2491-060X]{Max Gronke}
\affiliation{Centre for Astronomy of Heidelberg University, Astronomisches Rechen-Institut, M\"onchhofstr. 12-14, 69120 Heidelberg, Germany}
\email{max.gronke@uni-heidelberg.de}

\author[0000-0001-6278-032X]{Lukas J. Furtak}
\affiliation{Cosmic Frontier Center, The University of Texas at Austin, Austin, TX 78712, USA}
\affiliation{Department of Astronomy, The University of Texas at Austin, Austin, TX 78712, USA}
\email{furtak@utexas.edu}

\author[0000-0002-0112-5900]{Seok-Jun Chang}
\affiliation{Max-Planck-Institut f\"ur Astrophysik, Karl-Schwarzschild-Stra\ss e 1, D-85748 Garching b. M\"unchen, Germany}
\email{sjchang@MPA-Garching.MPG.DE}

\author[0000-0003-2871-127X]{Jorryt Matthee}
\affiliation{Institute of Science and Technology Austria (ISTA), Am Campus 1, 3400 Klosterneuburg, Austria}
\email{jorryt.matthee@ist.ac.at}

\author[0000-0003-1432-7744]{Lily Whitler}
\affiliation{Kavli Institute for Cosmology, University of Cambridge, Madingley Road, Cambridge, CB3 0HA, UK}
\affiliation{Cavendish Laboratory, University of Cambridge, JJ Thomson Avenue, Cambridge, CB3 0US, UK}
\email{lw851@cam.ac.uk}

\author[0000-0002-0350-4488]{Adi Zitrin}
\affiliation{Department of Physics, Ben-Gurion University of the Negev, P.O. Box 653, Be’er-Sheva 84105, Israel}
\email{adizitrin@gmail.com}

\author[0000-0003-4564-2771]{Ryan Endsley}
\affiliation{Department of Astronomy, The University of Texas at Austin, Austin, TX 78712, USA}
\affiliation{Cosmic Frontier Center, The University of Texas at Austin, Austin, TX 78712, USA}
\email{}

\author[0000-0001-5487-0392]{Viola Gelli}
\affiliation{Cosmic Dawn Center (DAWN)}
\affiliation{Niels Bohr Institute, University of Copenhagen, Jagtvej 128, 2200 Copenhagen N, Denmark}
\email{viola.gelli@nbi.ku.dk}

\author[0009-0003-9906-2745]{Tamojeet Roychowdhury}
\affiliation{Department of Astronomy, University of California, Berkeley, Berkeley, CA 94720, USA}
\email{tamojeet@berkeley.edu}

\author[0000-0002-9132-6561]{Peter Senchyna}
\affiliation{The Observatories of the Carnegie Institution for Science, 813 Santa Barbara Street, Pasadena, CA 91101, USA}
\email{psenchyna@carnegiescience.edu}

\author[0000-0001-8426-1141]{Michael W. Topping}
\affiliation{Steward Observatory, University of Arizona, 933 N Cherry Avenue, Tucson, AZ 85721, USA}
\email{}

\author[0009-0008-4707-6092]{Meng Zhang}
\affiliation{Tsung-Dao Lee Institute, Shanghai Jiao Tong University, 1 Lisuo Road, Shanghai 201210, People’s Republic of China}
\affiliation{School of Physics and Astronomy, Shanghai Jiao Tong University, 800 Dongchuan Road, Shanghai 200240, People’s Republic of China}
\affiliation{State Key Laboratory of Dark Matter Physics, Shanghai Jiao Tong University, 1 Lisuo Road, Shanghai 201210, People’s Republic of China}
\email{mengzhangastro@sjtu.edu.cn}

\begin{abstract}
Rest-frame ultraviolet (UV) spectra of Little Red Dots (LRDs) often show Ly$\alpha$ emission. 
Along with broad Balmer emission, LRDs are expected to produce broad Ly$\alpha$ emission. 
However, the large column density of neutral gas invoked to explain the Balmer break should significantly redshift and further broaden the Ly$\alpha$ line, making it challenging to detect without sensitive, moderate-resolution spectra.
We present ultra-deep (29 hours) G140M JWST/NIRSpec observations covering the rest-UV of two LRDs in Abell2744 from the SPURS Cycle 4 Large Program. 
One of our targets is Abell2744-QSO1, a gravitationally-lensed LRD at $z=7.04$ with faint UV emission (M$_{\rm{UV}}=-16.9$), and the other source (UNCOVER-2476) is newly-confirmed at $z=4.02$ with a very bright UV continuum (M$_{\rm{UV}}=-19.6$). 
We find that Abell2744-QSO1 has a broad Ly$\alpha$ profile, along with narrow C~{\small IV}, Fe~{\small II}~$\lambda1786$, and O~{\small I}~$\lambda1302$ emission. 
The Ly$\alpha$ profile suggests an origin similar to the broad H$\alpha$, but the line is considerably less redshifted than expected from existing dense gas models. 
We show that the line profile can be explained if the dense neutral gas is clumpy, allowing Ly$\alpha$ to escape by scattering off of the clump surfaces. 
We find that UNCOVER-2476 has narrow [Ne~{\small IV}] emission, indicating either a hard radiation field or shocks. 
We confirm two close neighbors with Ly$\alpha$ emission around Abell2744-QSO1, indicating it traces a dense environment that may have ionized its surrounding IGM. 
We suggest that LRDs may preferentially trace bubbles carved by their dense environments, contributing to the prevalence of Ly$\alpha$ in the population. 
\end{abstract}



\section{Introduction} \label{sec:intro}

The launch of JWST \citep{Gardner2023,Rigby2023} opened a new window on active galactic nuclei (AGNs) and the growth of supermassive black holes (SMBHs) in the high redshift universe. 
One of the biggest surprises has been the discovery of the Little Red Dots (LRDs), sources characterized by very red continua in the rest-frame optical, blue colors in the rest-frame ultraviolet (UV), and both broad hydrogen Balmer lines and narrow forbidden lines in the rest-frame optical \citep[e.g.,][]{Harikane2023,Greene2024,Matthee2024,Kocevski2025,Labbe2025}. 
Application of lower redshift virial relations to the broad line measurements has led to the suggestion that LRDs are associated with galaxies hosting black holes with masses of $\simeq10^6$ to $10^8\ M_{\odot}$ \citep[e.g.,][]{Greene2024,Lin2024,Maiolino2024b,Matthee2024,Kocevski2025}.
Simple estimates of the LRD host properties point to relatively low stellar masses ($\simeq10^8-10^9\ M_{\odot}$; e.g., \citealt{Maiolino2024b,Kocevski2025}), indicating that the LRDs may trace galaxies with ``overmassive'' black holes at high redshift.
Current measurements of the volume density of LRDs at $z\gtrsim 4$ reveal this may represent a key phase in the early growth of SMBHs \citep[e.g.,][]{Greene2024,Kokorev2024,Akins2025b,Kocevski2025,Labbe2025}. 

However, it has become clear in the last several years that LRD properties differ from those of most lower-redshift AGN. 
In particular, LRDs are undetected or only weakly detected in X-rays \citep[e.g.,][]{Ananna2024,Yue2024,Akins2025b,Maiolino2025a}, lack evidence for hot dust emission in the rest-frame near-infrared (NIR; e.g., \citealt{Perez-Gonzalez2024,Williams2024,Akins2025b}), are generally undetected at radio wavelengths \citep[e.g.,][]{Casey2025,Gloudemans2025,Setton2025,Xiao2025}, and show little variability \citep[e.g.,][]{Furtak2025,Kokubo2025,Tee2025,Zhang2025}. 
They also often do not show evidence for strong high ionization lines (N~{\small V}, C~{\small IV}, He~{\small II}, [Ne~{\small IV}], [Ne~{\small V}]) typically seen in AGN \citep[e.g.,][]{Lambrides2024,Wang2025b}, although there are LRDs which do clearly have hard radiation fields \citep{Labbe2024,Akins2025a,Tang2025,Treiber2025}. 

One potential explanation for the unique properties of LRDs is related to a large covering fraction of extremely dense gas. 
The existence of dense absorbing gas surrounding LRDs was first indicated by narrow H$\beta$ and H$\alpha$ absorption features superimposed on the broad Balmer emission lines \citep[e.g.,][]{Lin2024,Matthee2024,Ji2025,Kocevski2025,Taylor2025,DEugenio2026}. 
These can be explained via the presence of neutral gas densities of $\gtrsim 10^9$~cm$^{-3}$ that are capable of collisionally exciting the $n=2$ atomic level of hydrogen \citep[e.g.][]{Juodzbalis2024,Inayoshi2025,Ji2025,Naidu2025}. 
If such dense gas is present, there should also be strong continuum absorption at the Balmer limit. 
As the first prism spectra of LRDs emerged, it became clear that many indeed exhibit strong Balmer breaks \citep[e.g.,][]{Furtak2024,Juodzbalis2024,Labbe2024,Wang2024,deGraaff2025b,Naidu2025}, confirming the impact that dense gas has in shaping the LRD spectrum.

If the AGN is fully enshrouded and column densities are large enough to thermalize the radiation, a quasi-blackbody spectrum may be formed, with effective temperature of $T_{\rm eff}\simeq 5000$~K that is set by hydrogen opacity \citep[e.g.,][]{deGraaff2025c,Inayoshi2026,Kido2025,Liu2025,Umeda2026b}. 
In this case, the red optical continuum of LRDs corresponds to the Wien tail of the spectrum, and the absence of X-rays and hot dust emission naturally follows from the Compton thick column densities. 
It has been argued that the Balmer lines may be broadened via electron scattering as they traverse ionized regions within the column of dense gas \citep[e.g.,][]{Chang2026,Rusakov2026,Sneppen2026,Torralba2026b}. 
In this case, the intrinsic line widths may be significantly narrower than measured, and potentially decoupled from the mass, which in turn may suggest that many LRDs may have lower black hole masses than inferred via the virial relations (e.g., \citealt{DEugenio2025a,Kokorev2025,Naidu2025,Rusakov2026}; c.f., \citealt{Brazzini2025}). 

While the dense-gas cocoon model provides a plausible framework for many properties of LRDs, the picture is still not universally accepted and key questions remain, with several non-AGN scenarios also proposed \citep[e.g.,][]{Zwick2025,Chisholm2026,Nandal2026}.
New constraints on the dense gas coverage are required to stress-test the emerging picture. 
Ly$\alpha$ is well known to provide a sensitive probe of dense neutral gas \citep[e.g.,][]{Neufeld1990,Dijkstra2017}. 
If the neutral hydrogen column densities are as large as required to explain the Balmer break and line absorption ($\gtrsim10^{24}$~cm$^{-2}$; e.g., \citealt{Juodzbalis2024,Inayoshi2025,Ji2025}), the Ly$\alpha$ line will undergo substantial resonant scattering as it traverses the dense gas, significantly altering the line profile with respect to the broad Balmer lines \citep{Adams1972,Neufeld1990,Verhamme2006}. 
While Ly$\alpha$ has been detected in many LRDs \citep[e.g.,][]{Kokorev2023,Furtak2024,Asada2026}, very few have been observed at the spectral resolution and depth required to test the predictions of the dense gas cocoon picture \citep{Morishita2026,Torralba2026a}.

The high ionization lines provide another probe of the dense gas cocoon picture. 
If the LRD continuum is associated with a photosphere with $T_{\rm eff}\simeq5000$~K, we should not expect hard radiation to be transmitted to the narrow-line emitting region. 
However, as noted above, several very high ionization emission lines have been detected in LRD prism and shallow grating spectra \citep{Labbe2024,Akins2025a,Lambrides2025,Tang2025,Treiber2025}. 
The origin of these high ionization lines is not known. 
Stellar populations can produce hard photons capable of powering strong C~{\small IV} and He~{\small II}, but emission lines that probe above the He$^+$-ionizing edge ($>54$~eV) are more likely to be linked to accretion onto a supermassive black hole \citep[e.g.,][]{Feltre2016}. 
The presence of very high ionization lines may point to a non-uniform coverage of dense neutral gas, suggesting that at least some LRDs may not be fully enshrouded in gas cocoons. 
Unfortunately little is known about how commonly LRDs exhibit emission from the most prominent high ionization lines (i.e., N~{\small V}, C~{\small IV}, He~{\small II}, [Ne~{\small IV}], [Ne~{\small V}]) given the limited sensitivity of the low resolution prism (see \citealt{Tang2025} for discussion). 
In cases where permitted high ionization (i.e., C~{\small IV}, He~{\small II}) lines have been detected in prism spectra, the resolution has obscured whether the lines originate in the broad line region (BLR), or whether they reflect the escape of hard photons to the narrow line region (NLR).

Progress requires deep observations of LRDs in the rest-frame UV using higher spectral resolution than has been obtained to-date.
In this paper, we present the first results from our Cycle 4 program the SPectroscopic Ultra-deep Reionization-era Survey (SPURS; GO 9214, PIs: C. Mason, D. Stark). 
SPURS provides ultra deep (29~hours) median-resolution ($R\simeq1000$) grating spectra in G140M. 
One of our goals is to obtain robust constraints on the Ly$\alpha$ and very high ionization UV emission lines (N~{\small V}, C~{\small IV}, He~{\small II}, [Ne~{\small IV}], [Ne~{\small V}]) in the rest-frame UV spectra of LRDs and broad-line AGNs at high redshift. 
Here we focus on the observations of two LRDs behind the lensing galaxy cluster Abell 2744: Abell2744-QSO1 and UNCOVER-2476. 
Abell2744-QSO1 was first identified as an extremely red and compact object in \citet{Furtak2023a} from the Ultra-deep NIRSpec and NIRCam ObserVations before the Epoch of Reionization (UNCOVER) survey (GO 2561, PIs: I. Labb\'{e}, R. Bezanson; \citealt{Bezanson2024}). 
Subsequent NIRSpec \citep{Jakobsen2022,Boker2023} low-resolution ($R\sim100$) prism spectroscopy revealed a strong Ly$\alpha$ emission, broad H$\beta$ and H$\alpha$ emission, and strong Balmer break feature at $z=7.04$ \citep{Furtak2024}. 
High-resolution ($R\simeq2700$) G395H integral field unit (IFU) spectroscopy further revealed H$\beta$ and H$\alpha$ absorption lines in Abell2744-QSO1 \citep{DEugenio2025a,Ji2025}. 
UNCOVER-2476 was photometrically identified in \citet{Labbe2025} at $z_{\rm phot}=4.56$. 
Rest-frame UV grating spectroscopy has not been obtained for either of these two LRDs, and UNCOVER-2476 has yet to be observed with any spectrograph.

The organization of this paper is as follows. 
In Section~\ref{sec:obs}, we describe SPURS observations of Abell2744-QSO1 and UNCOVER-2476, the basic physical properties of these two LRDs, as well as two newly identified galaxies nearby Abell2744-QSO1. 
We then characterize the rest-frame UV spectroscopic properties of Abell2744-QSO1 and UNCOVER-2476 from SPURS in Section~\ref{sec:lrd_uv_spec}. 
We investigate the H~{\small I} gas properties inferred from the Ly$\alpha$ emission of Abell2744-QSO1 in Section~\ref{sec:lya_model}, and discuss the implications for the environment and structure of LRDs in Section~\ref{sec:discussion}. 
Finally, we summarize our conclusions in Section~\ref{sec:summary}. 
Throughout the paper we adopt a $\Lambda$-dominated, flat universe with $\Omega_{\Lambda}=0.7$, $\Omega_{\rm{M}}=0.3$, and $H_0=70$~km~s$^{-1}$~Mpc$^{-1}$. 
All magnitudes are quoted in the AB system \citep{Oke1983} and all equivalent widths (EWs) are quoted in the rest frame.

\section{Observations and Data Analysis} \label{sec:obs}

\subsection{Spectroscopic Observations} \label{sec:spec_obs}


\begin{figure}
\includegraphics[width=\linewidth]{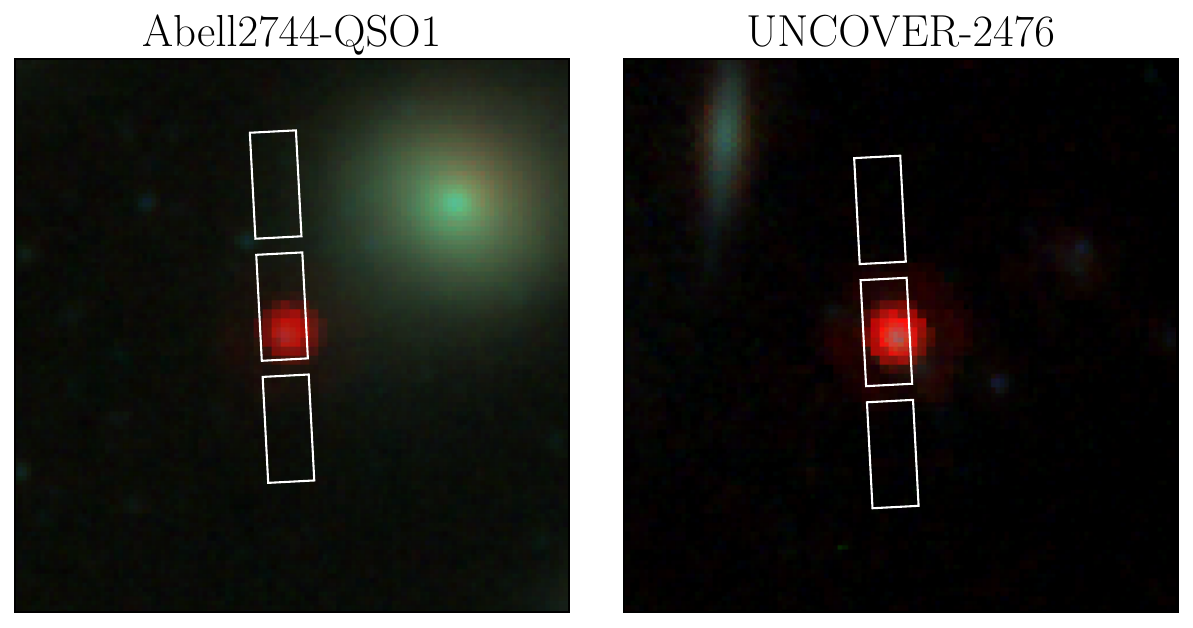}
\caption{JWST/NIRCam RGB (F444W/F200W/F115W) images ($2.4''\times2.4''$) of Abell2744-QSO1 (left) and UNCOVER-2476 (right). We overplot our NIRSpec shutters (white) in SPURS observations. Abell2744-QSO1 is triply imaged by Abell 2744 and we target its the mostly strongly magnified image (`image B'' in \citealt{Furtak2023a}).}
\label{fig:image}
\end{figure}

We describe the JWST/NIRSpec spectroscopic data of Abell2744-QSO1 (RA $=3.583540$, Dec $=-30.396677$)\footnote{Abell2744-QSO1 is triply imaged by the galaxy cluster Abell 2744, and we target the mostly strongly lensed image (``image B'' in \citealt{Furtak2023a}) in SPURS observations.} and UNCOVER-2476 (RA $=3.610205$, Dec $=-30.421001$) that were obtained as part of SPURS program (Figure~\ref{fig:image}). 
The spectra were obtained using NIRSpec in multi-object spectroscopy mode in 2025 November. 
We observed one microshutter assembly (MSA; \citealt{Ferruit2022}) mask configuration in the Abell 2744 field, with Abell2744-QSO1 as one of the primary targets. 
We briefly summarize the SPURS NIRSpec observations of the two LRDs below. 
A full description of the SPURS program will be presented in a future paper.

The NIRSpec observations were conducted using the medium-resolution ($R\simeq1000$, corresponding to $\simeq300$~km~s$^{-1}$ per resolution element) grating/filter pairs G140M/F100LP, G235M/F170LP, and G395M/F290LP. 
We used the three-shutter nod pattern for dithering, which is appropriate for compact high redshift targets. 
The total on-target integration time is 29.2~hours, 7.9~hours, and 2.9~hours for G140M, G235M, and G395M, respectively.

We reduced the 2D G140M and G235M + G395M spectra separately. 
We first reduced the G140M spectra following the approaches described in \citet{Topping2025} which are based on the standard JWST data reduction pipeline\footnote{\url{https://github.com/spacetelescope/jwst}} \citep{Bushouse2024}. 
This is customized to produce a pixel size of $3.5$~\AA\ in the spectral direction for G140M, which is optimal for characterizing the line profiles of Ly$\alpha$ and UV emission for the two LRDs. 
The G235M and G395M spectra were reduced following the procedures and setup described in \citet{deGraaff2025a,Heintz2025} using the latest version of \texttt{msaexp}\footnote{\url{https://github.com/gbrammer/msaexp}} package \citep{Brammer2023}, which is also based on the standard pipeline. 
Our reduction with \texttt{msaexp} results in a slightly coarser dispersion ($6$~\AA\ in G140M), but its extended wavelength extraction allows us to achieve longer wavelengths than the normal coverage of each grating. 
The reduced G235M and G395M spectra cover $1.63-3.02\ \mu$m and $2.80-5.46\ \mu$m, respectively. 
This allows us to cover the H$\alpha$ emission of Abell2744-QSO1 ($5.28\ \mu$m) which will be discussed in the following sections. 
We confirm that the emission line fluxes and widths measured from the spectra reduced by these two methods are identical.
We assumed a point source pathloss correction, motivated by the compact morphology of LRDs (Figure~\ref{fig:image}). 

The 1D spectra of the two LRDs were extracted from the reduced 2D spectra using a boxcar extraction with an aperture of 5 pixels, corresponding to $\sim0.5''$ in the spatial direction. 
For G140M spectra, the median $3\sigma$ limiting flux for an unresolved emission line (in spectral direction) of a point source is $1.0\times10^{-19}$~erg~s$^{-1}$~cm$^{-2}$. 
This allows us to detect very weak emission lines in the rest-frame UV for Abell2744-QSO1 (EW $=2.9$~\AA) and UNCOVER-2476 (EW $=0.9$~\AA). 
The median $3\sigma$ limiting flux of G235M spectra is $1.9\times10^{-19}$~erg~s$^{-1}$~cm$^{-2}$, corresponding to a limiting EW of $9.9$~\AA\ for Abell2744-QSO1 and $2.6$~\AA\ for UNCOVER-2476. 
While most of our focus will be on the two bluer gratings, we also note that the G395M spectra reach a median $3\sigma$ limiting flux of $2.0\times10^{-19}$~erg~s$^{-1}$~cm$^{-2}$. 


\begin{figure*}
\includegraphics[width=\linewidth]{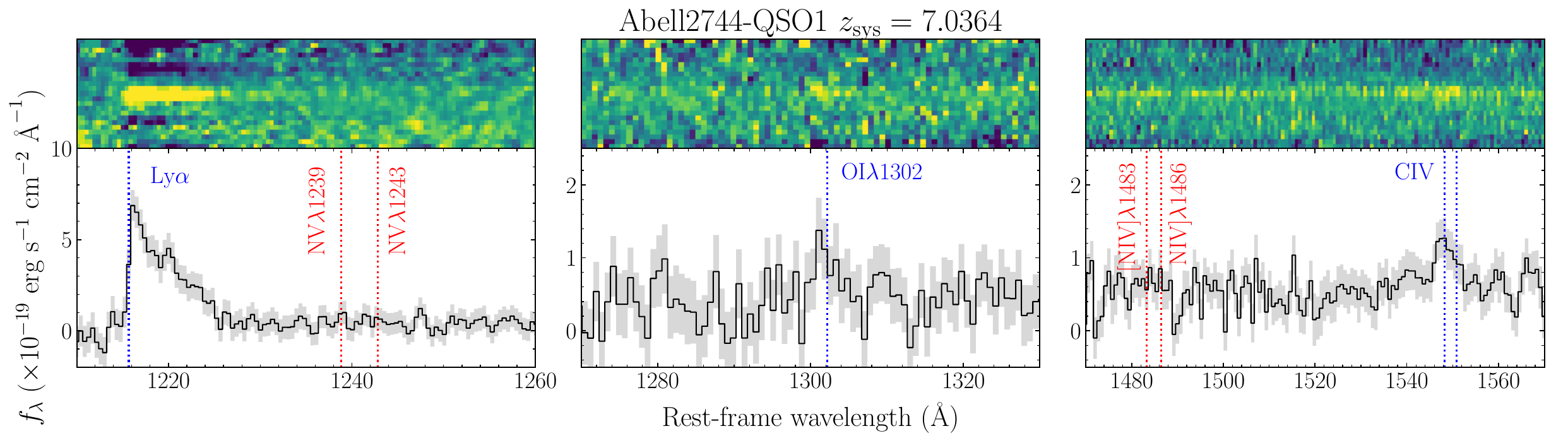}
\includegraphics[width=\linewidth]{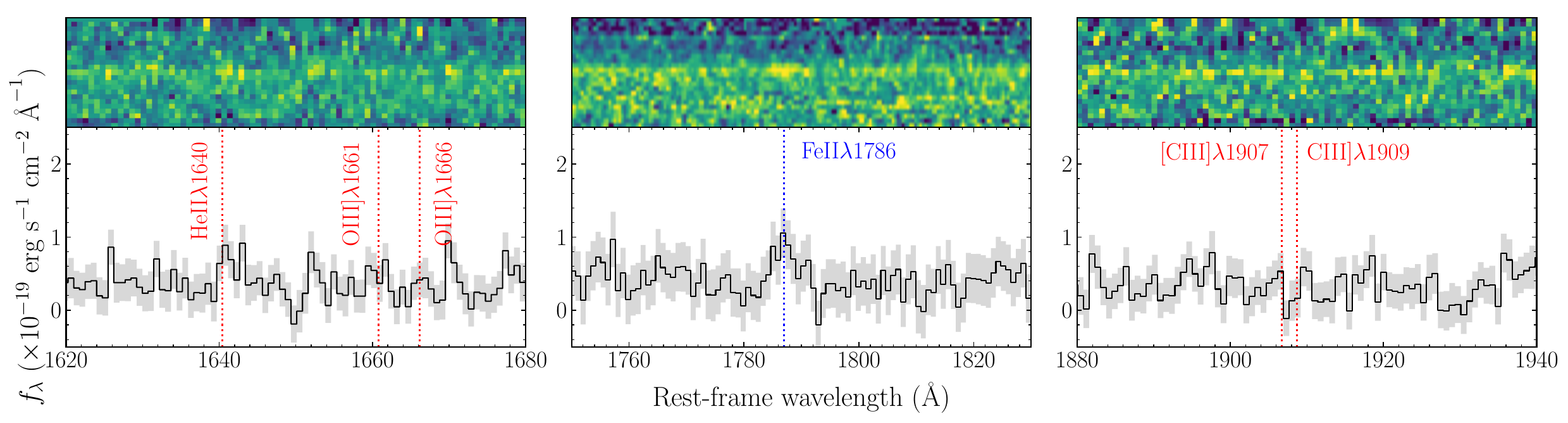}
\includegraphics[width=\linewidth]{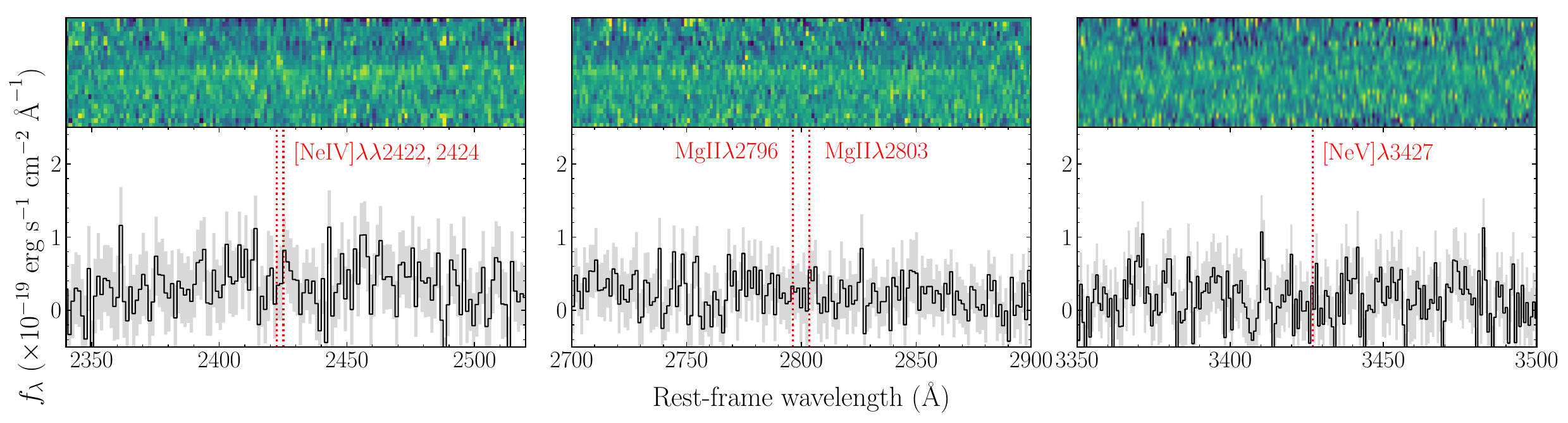}
\caption{Deep JWST/NIRSpec rest-frame UV (G140M/F100LP and part of G235M/F170LP) spectra of Abell2744-QSO1 obtained from SPURS program. In each panel, we show the 2D spectrum at the top and the 1D at the bottom. The rest-frame UV continuum is marginally detected. We detect the Ly$\alpha$, O~{\scriptsize I}~$\lambda1302$, C~{\scriptsize IV}~$\lambda\lambda1548,1551$, and Fe~{\scriptsize II}~$\lambda1786$ emission lines, overplotting their expected positions from the systemic redshift ($z_{\rm sys}=7.0364$) with blue dotted lines. We also plot the expected positions of non-detections (N~{\scriptsize V}, N~{\scriptsize IV}], He~{\scriptsize II}, O~{\scriptsize III}], C~{\scriptsize III}], [Ne~{\scriptsize IV}], Mg~{\scriptsize II}, [Ne~{\scriptsize V}]) with red dotted lines.}
\label{fig:QSO1_uv_spec}
\end{figure*}

The two LRD spectra have numerous emission lines, which we characterize as follows. 
For a line detected with S/N $>5$, we measure the line flux, centroid, width, and EW by fitting the line profile and nearby continuum with a Gaussian function.
In case of emission lines that are close in wavelength or an emission line that shows a complex profile, we fit the line profile with multiple Gaussians simultaneously. 
If an emission line is detected with lower S/N ($<5$), we compute the line fluxes using direct integration. 
Recent studies have found that the broad Balmer lines of LRDs are better fitted by exponential wings or double-Gaussian profiles when the S/N is high enough \citep[e.g.,][]{DEugenio2025a,DEugenio2025b,Kokorev2025,Matthee2026,Rusakov2026}. 
For the SPURS spectra of the two LRDs, we note that the line fluxes, EWs, and widths inferred from Gaussian fitting are similar to that inferred from exponential fitting. 
Therefore, we choose to use results from Gaussian fitting in this work, leaving a more detailed analysis of the broad Balmer line profiles to a future paper.
Finally, the uncertainties of line fluxes and EWs are evaluated as follows. 
We resample the flux densities of each spectrum $1000$ times by taking the observed value as the mean and the error as the standard deviation. 
Then we compute the line fluxes and EWs from the resampled spectra of each source using the same approach described above. 
We take the standard deviation of these measurements as the uncertainty. 
The line fluxes and uncertainties reported throughout the paper are not corrected for gravitational magnification.

\subsection{Physical Properties of LRDs in SPURS} \label{sec:lrd_properties}

\begin{figure*}
\includegraphics[width=\linewidth]{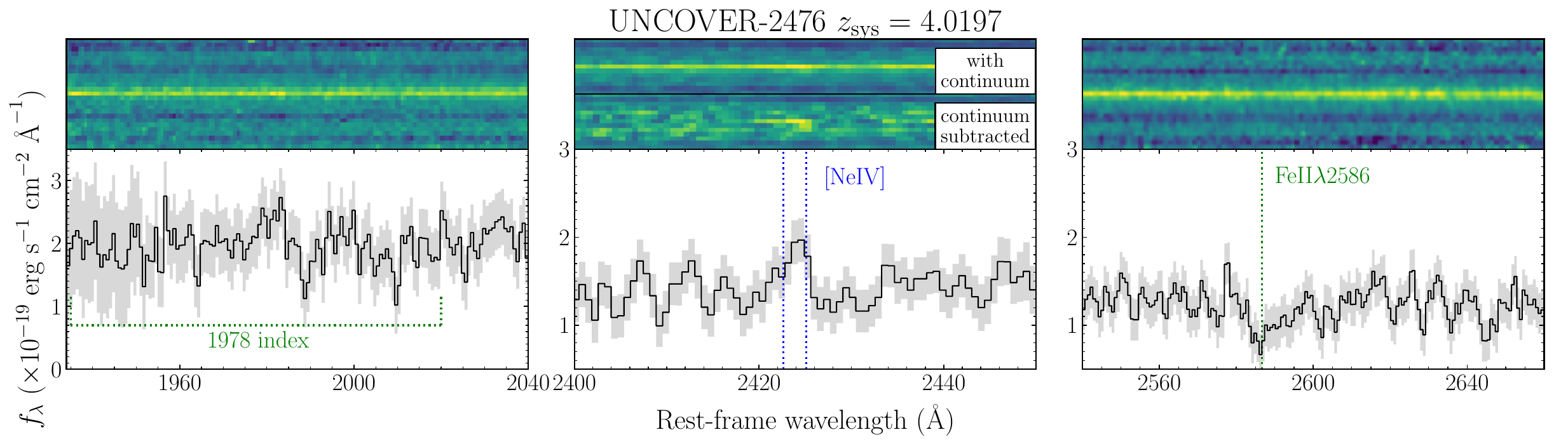}
\includegraphics[width=\linewidth]{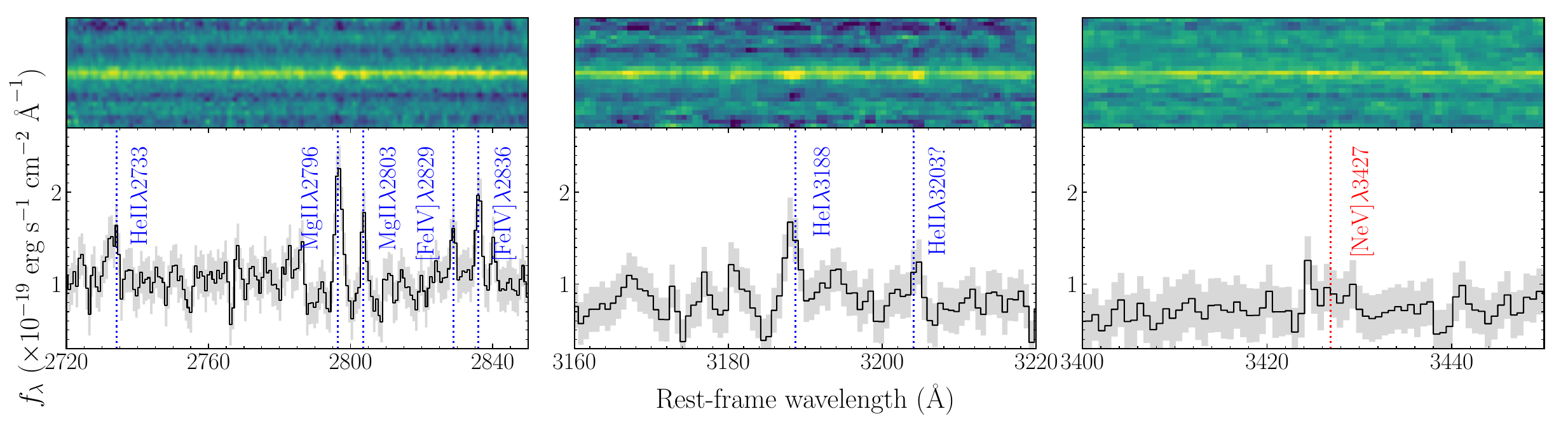}
\caption{Deep JWST/NIRSpec rest-frame UV (G140M/F100LP) spectrum of UNCOVER-2476 obtained from SPURS program, shown in the same way as Figure~\ref{fig:QSO1_uv_spec}. The rest-frame UV continuum is clearly detected. We also detect [Ne~{\scriptsize IV}]~$\lambda2422,2424$, He~{\scriptsize II}~$\lambda2733$, Mg~{\scriptsize II}~$\lambda2796,2803$, [Fe~{\scriptsize IV}]~$\lambda2829,2836$, He~{\scriptsize I}~$\lambda3188$ emission lines (blue), and tentatively detect He~{\scriptsize II}~$\lambda3203$. For [Ne~{\scriptsize IV}] (top middle panel), we present both the original 2D spectrum with continuum trace and the continuum-subtracted 2D spectrum, the later of which show the [Ne~{\scriptsize IV}] emission more clearly. [Ne~{\scriptsize V}]~$\lambda3427$ is not detected (red dotted line). We mark the $1978$ index photospheric absorption and the Fe~{\scriptsize II}~$\lambda2586$ absorption line with the green dashed lines.}
\label{fig:2476_uv_spec}
\end{figure*}

We briefly describe the basic physical properties of Abell2744-QSO1 and UNCOVER-2476, before discussing their deep rest-frame UV (G140M and part of G235M) spectra (Figure~\ref{fig:QSO1_uv_spec} and \ref{fig:2476_uv_spec}) in the following sections. 
Since the primary goal of this paper is the rest-frame UV, we defer a complete analysis of the SPURS rest-frame optical (G395M and part of G235M) spectra to a future paper. 
In the case of UNCOVER-2476, we describe more results from the rest-frame optical spectrum since this source had not been confirmed spectroscopically previously. 
In contrast, Abell2744-QSO1 has been studied with deep rest-frame optical spectroscopy \citep{DEugenio2025a,Ji2025,Juodzbalis2025,Maiolino2025b}, so here we primarily review basic properties from the literature, noting consistency with our new rest-frame optical spectrum. 


\subsubsection{Abell2744-QSO1} \label{sec:QSO1_info}

\begin{figure}
\includegraphics[width=\linewidth]{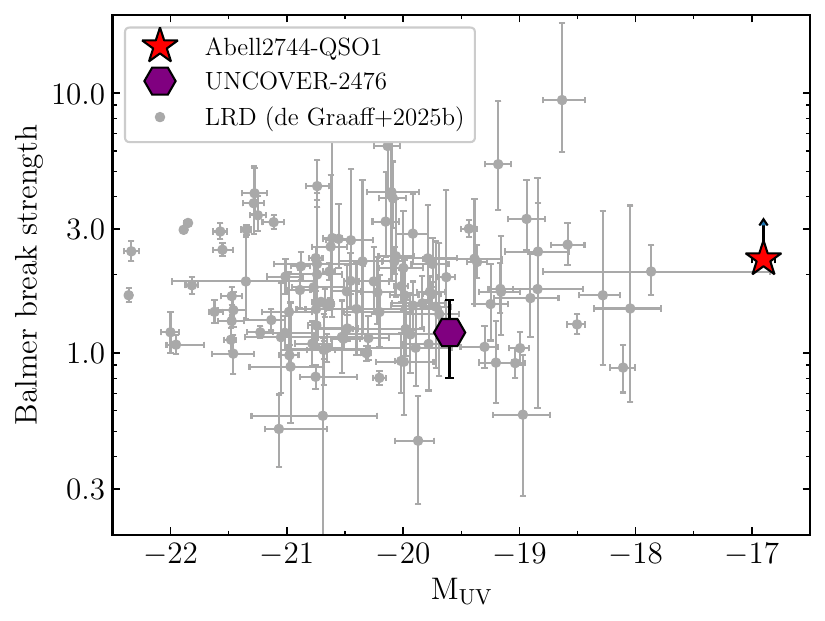}
\caption{M$_{\rm UV}$ versus Balmer break strength (defined by $f_{\nu,4050}/f_{\nu,3640}$ ratio) of the full LRD population. Abell2744-QSO1 is shown as the red star, with extremely faint rest-frame UV continuum and large Balmer break strength. UNCOVER-2476 is shown as the purple hexagon, with bright UV continuum and relatively small Balmer break strength. We overplot the $z\sim2-9$ LRD data from \citet{deGraaff2025c} as grey circles.}
\label{fig:muv_bb}
\end{figure}

Abell2744-QSO1 is extremely faint in rest-frame UV continuum (M$_{\rm UV}=-16.9$; \citealt{Furtak2023a}), with the lowest UV luminosity among the LRD population at high redshift (Figure~\ref{fig:muv_bb}). 
The image we targeted is highly magnified, with a magnification factor of $\mu=7.2$ derived using the \citet{Furtak2023b} lens models, which are updated with UNCOVER spectroscopic redshifts in \citet{Price2025}.
R100 prism and R2700 G395H IFU observations show broad H$\beta$ and H$\alpha$ emission (full width at half maximum FWHM $\simeq2700$~km~s$^{-1}$) with narrow [O~{\small III}]~$\lambda5007$ line \citep{Furtak2024,DEugenio2025a,Ji2025}. 
The [O~{\small III}]~$\lambda5007$ is very weak relative to the narrow H$\beta$ emission, likely indicating a very low metallicity for the narrow line emitting gas. 
\citet{Maiolino2025b} estimate the metallicity may be as low as $\simeq4\times10^{-3}\ Z_{\odot}$ based on the calibrations for [O~{\small III}]/H$\beta$ \citep{Laseter2024,Sanders2024}. 
The prism spectrum also revealed a strong Balmer break feature, with a strength ($f_{\nu,4050}/f_{\nu,3640}>2.3$; \citealt{Furtak2024,DEugenio2025a}) that is among the upper $15\%$ of the values observed in LRDs (Figure~\ref{fig:muv_bb}; \citealt{deGraaff2025c}). 
The G395H IFU spectrum further revealed strong absorption lines in both H$\beta$ (EW $=-5.5$~\AA; \citealt{Ji2025}) and H$\alpha$ (EW $=-30$~\AA; \citealt{DEugenio2025a}). 
Recent studies have suggested that the strong Balmer break and Balmer line absorption can be described by AGN emission absorbed by very dense gas \citep[e.g.,][]{Inayoshi2025,Ji2025,Naidu2025}. 
We will test this picture with our deep R1000 spectroscopy at rest-frame UV in the following sections. 

The SPURS G395M spectrum (covering rest-frame $3500-6800$~\AA) of Abell2744-QSO1 confirms the presence of broad H$\beta$ and H$\alpha$ emission with a weak, narrow [O~{\small III}]~$\lambda5007$ line (Figure~\ref{fig:QSO1_opt_spec}), consistent with earlier prism and G395H IFU observations. 
We derive a systemic redshift of $z_{\rm sys}=7.0364$ from the narrow forbidden [O~{\small III}]~$\lambda5007$ line in the G395M spectrum, in good agreement with G395H IFU measurement ($z=7.0367$; \citealt{DEugenio2025a}). 
We will adopt this systemic redshift for Abell2744-QSO1 throughout the paper. 
We also confirm that the broad H$\alpha$ line width in our G395M spectrum (FWHM$_{\rm H\alpha,broad}=2653\pm345$~km~s$^{-1}$) is consistent with G395H IFU measurements.
In Section~\ref{sec:QSO1_uv_spec}, we will compare the rest-frame UV line profiles to the broad H$\alpha$. 
We report the rest-frame optical emission line fluxes, EWs, and widths measured from our G395M spectrum of Abell2744-QSO1 in Table~\ref{tab:QSO1_opt_line}.


\subsubsection{UNCOVER-2476} \label{sec:2476_info}

The SPURS observations have obtained the first spectrum of UNCOVER-2476. 
The G235M and G395M spectra (covering rest-frame $3250-10950$~\AA) reveal a suite of rest-frame optical to NIR emission lines including broad H$\beta$ and H$\alpha$ emission with narrow forbidden lines (Figure~\ref{fig:2476_opt_nir_spec}), as is often seen in broad line AGN and LRDs. 
Using the strong narrow [O~{\small III}] doublet, we derive a systemic redshift of $z_{\rm sys}=4.0197$. 
We measure a FWHM of $1703\pm28$~km~s$^{-1}$ ($2123\pm496$~km~s$^{-1}$) for the broad H$\alpha$ (H$\beta$) line, similar to the typical broad line width seen in LRDs (FWHM $\simeq2000$~km~s$^{-1}$; e.g., \citealt{Greene2024,Lin2024,Maiolino2024b,Matthee2024,Hviding2025,Kocevski2025}). 
We report the rest-frame optical to NIR emission line measurements of UNCOVER-2476 in Table~\ref{tab:2476_opt_nir_line}. 
In the following, we aim to briefly characterize the rest-frame optical spectroscopic properties of UNCOVER-2476. 
We leave a more detailed description of the rest-frame optical spectrum in Appendix~\ref{sec:2476_opt_nir_spec}.

UNCOVER-2476 is much brighter in rest-frame UV continuum (M$_{\rm UV}=-19.6$) relative to Abell2744-QSO1, comparable to the typical M$_{\rm UV}$ ($\simeq-20$) of the LRD population (Figure~\ref{fig:muv_bb}; \citealt{deGraaff2025c}). 
Its image is moderately magnified, with a magnification factor of $\mu=1.9$. 
The G235M spectrum reveals strong narrow emission lines in rest-frame optical. 
We measure a large [O~{\small III}]~$\lambda5007$ EW of $673\pm5$~\AA. 
This is not only much larger than the EW of Abell2744-QSO1 ($5.0\pm1.5$~\AA), but also among the upper $10\%$ of the EWs observed in LRDs \citep{deGraaff2025c}. 
We also detect a strong (EW $=-6.6\pm0.9$~\AA), narrow (FWHM $=184\pm92$~km~s$^{-1}$) H$\alpha$ absorption line that is blueshifted ($-257\pm92$~km~s$^{-1}$) from the line center (Appendix~\ref{sec:2476_opt_nir_spec}). 
Recent studies have pointed to a connection between Balmer absorption feature and Balmer break in LRDs, suggesting the absorption by dense gas as a potential explanation \citep[e.g.,][]{deGraaff2025b,Inayoshi2025,Ji2025,Naidu2025}. 
However, we do not detect a strong Balmer break feature for UNCOVER-2476. 
From our G235M spectrum, we measure a Balmer break strength that is close to unity ($f_{\nu,4050}/f_{\nu,3640}=1.2\pm0.4$; Figure~\ref{fig:muv_bb}). 
This is much weaker than the Balmer breaks of many well-studied LRDs with Balmer absorption lines ($f_{\nu,4050}/f_{\nu,3640}\simeq3-10$; \citealt{Furtak2024,Labbe2024,Wang2024,Wang2025a,deGraaff2025b,Naidu2025}). 
We will further characterize the properties of UNCOVER-2476 with deep rest-frame UV spectrum in Section~\ref{sec:2476_uv_spec}.

\subsection{Two New $z\simeq7.04$ Galaxies Near Abell2744-QSO1} \label{sec:z7_lae}


\begin{figure}
\includegraphics[width=\linewidth]{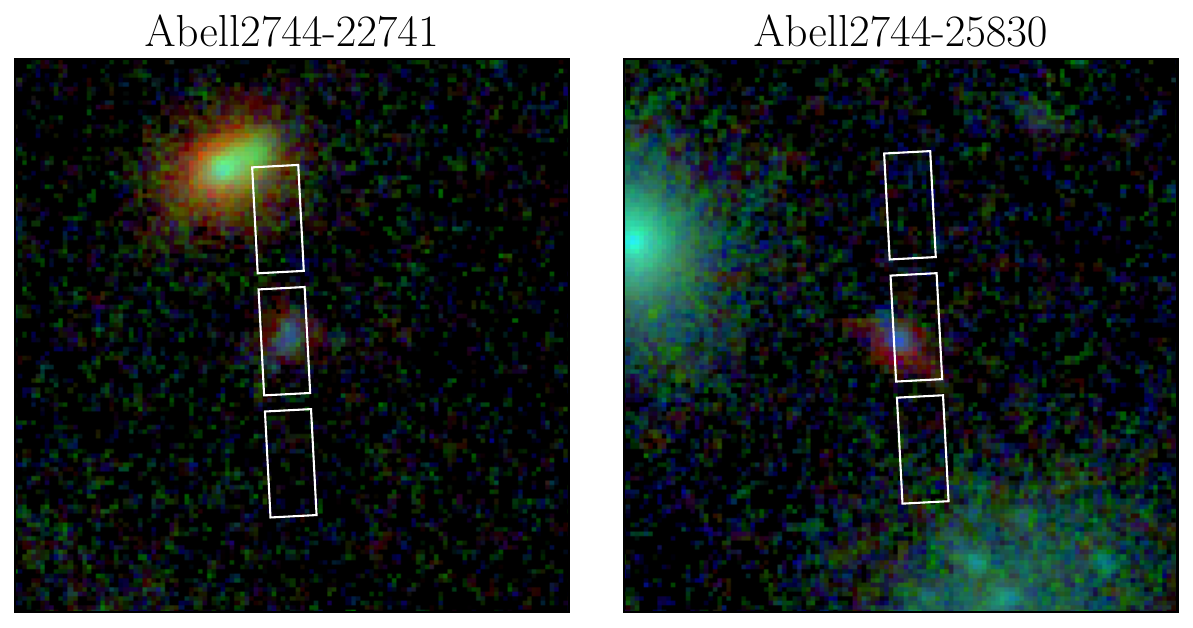}
\caption{JWST/NIRCam RGB (F444W/F200W/F115W) images ($2.4''\times2.4''$) of the two new $z\simeq7.04$ galaxies identified from SPURS: Abell2744-22741 (left) and Abell2744-25830 (right). We overplot our NIRSpec shutters (white) in SPURS observations.}
\label{fig:image_LAE}
\end{figure}


\begin{figure*}
\centering
\includegraphics[width=0.49\linewidth]{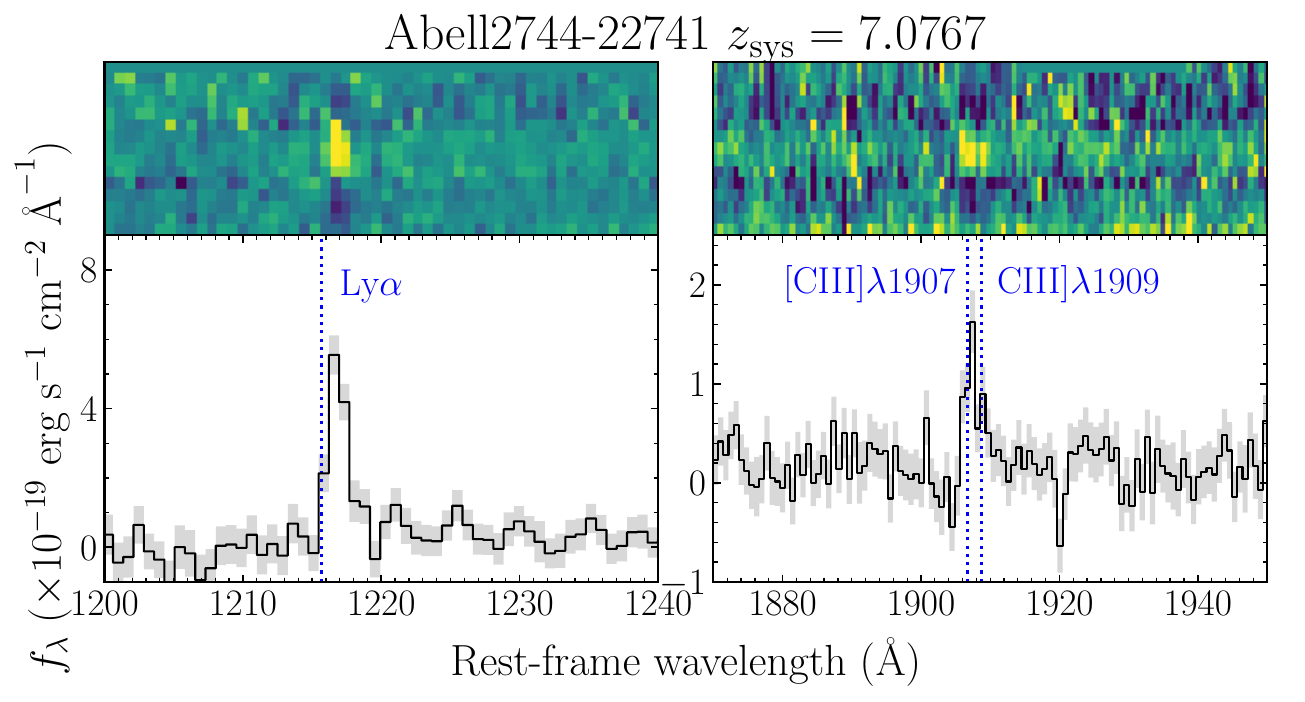}
\includegraphics[width=0.5\linewidth]{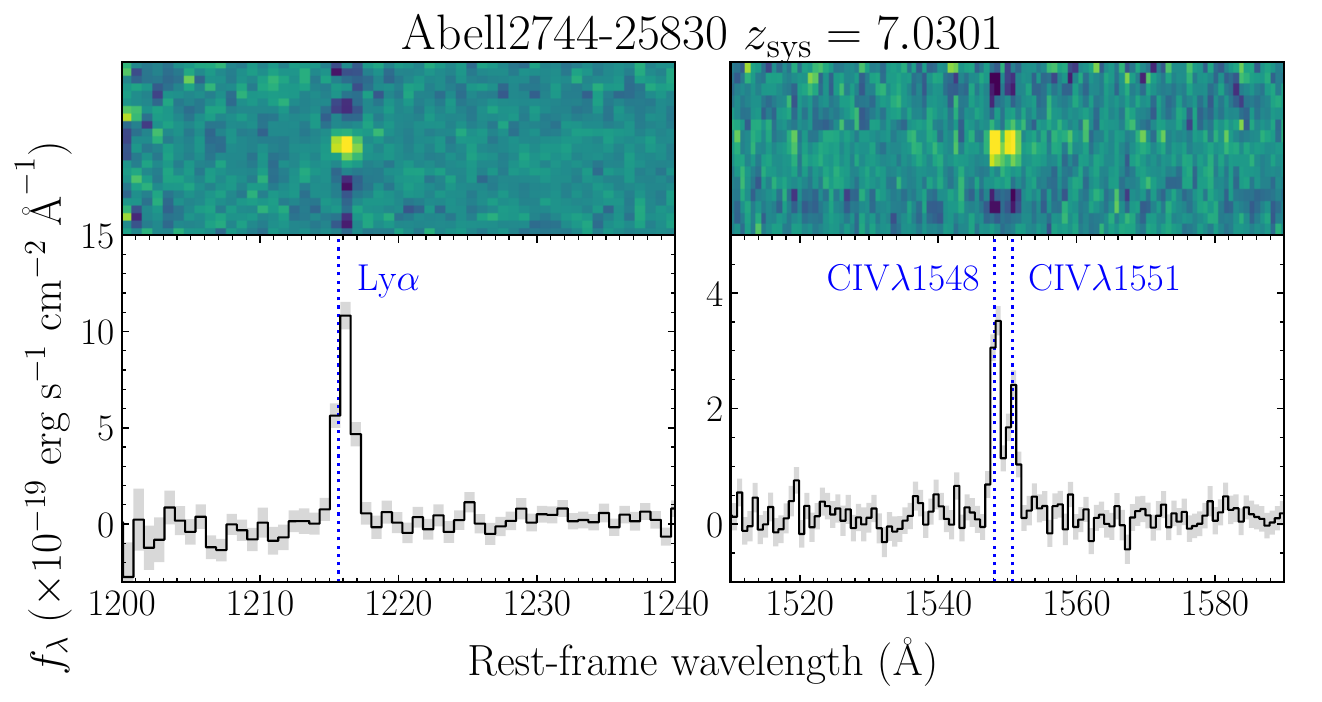}
\includegraphics[width=0.49\linewidth]{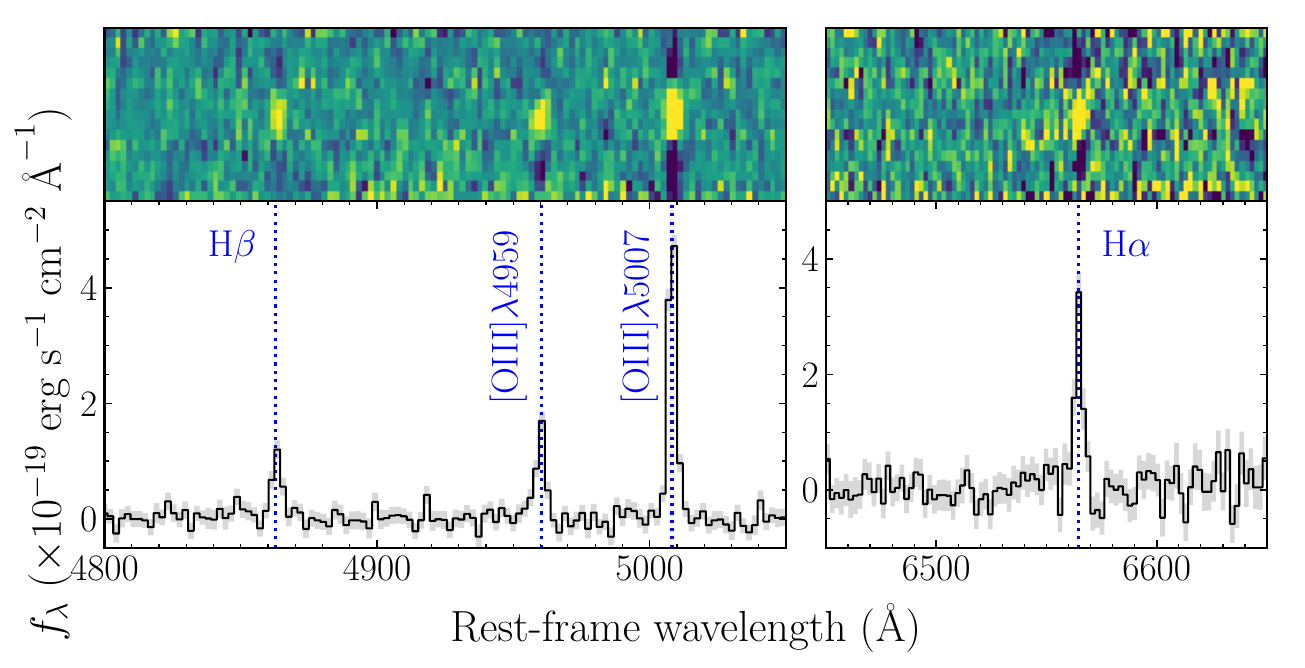}
\includegraphics[width=0.5\linewidth]{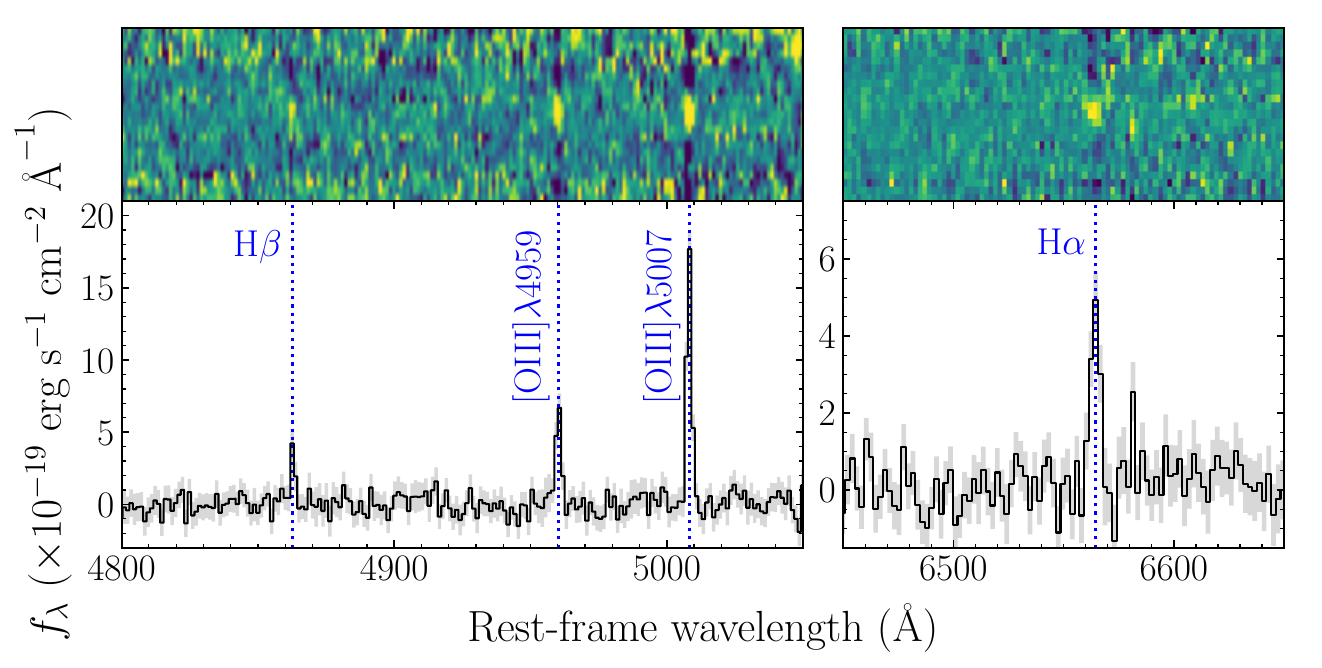}
\caption{SPURS spectra of the two galaxies nearby Abell2744-QSO1: Abell2744-22741 (left four panels) and Abell2744-25830 (right four panels). For each object, we show the rest-frame UV (G140M) spectrum at the top and the rest-frame optical (G395M) at the bottom. Ly$\alpha$ and strong rest-frame optical emission lines (H$\beta$, [O~{\scriptsize III}], H$\alpha$) are detected in both systems. We additionally detect C~{\scriptsize III}] in Abell2744-22741 and C~{\scriptsize IV} in Abell2744-25830. Spectra are shown in the same way as Figure~\ref{fig:QSO1_uv_spec}.} 
\label{fig:z7_lae_spec}
\end{figure*}

From SPURS observations on the Abell 2744 field, we also newly identify two galaxies at redshifts that are close to Abell2744-QSO1 ($z\simeq7.04$): Abell2744-22741 (RA $=3.625925$, Dec $=-30.393111$) and Abell2744-25830 (RA $=3.598496$, Dec $=-30.412591$). We present these in this paper, owing to implications for the Ly$\alpha$ visibility of Abell2744-QSO1.
The photometry of these two sources was initially characterized in \citet{Endsley2025}, identifying them as $z\sim7$ candidates. 
The SPURS observations have obtained the first spectra of these two sources (Figure~\ref{fig:image_LAE}). 
We briefly describe the spectroscopic properties of these two systems in this subsection, with the goal of understanding the galaxy environment associated with Abell2744-QSO1. 

Abell2744-22741 is more UV-luminous (M$_{\rm UV}=-18.8$) than Abell2744-QSO1.
We present the SPURS spectra of Abell2744-22741 in the left panels of Figure~\ref{fig:z7_lae_spec}. 
This galaxy is $0.68$~physical~Mpc (pMpc) away from Abell2744-QSO1 in the source plane after correcting for gravitational deflection with the updated \citet{Furtak2023b} lens models.
The G395M (rest-frame optical) spectrum reveals a suite of emission lines (e.g., H$\beta$, [O~{\small III}]~$\lambda4959$, [O~{\small III}]~$\lambda5007$, H$\alpha$), identifying that this galaxy is at $z_{\rm sys}=7.0767$. 
The rest-frame optical emission lines are strong, with an [O~{\small III}]+H$\beta$ EW ($843\pm42$~\AA) that is larger than the median EW of $z\sim7$ galaxies with similar UV luminosities ($\simeq500-600$~\AA\ at M$_{\rm UV}\simeq-18.6$; \citealt{Endsley2024,Begley2025}). 
This likely suggests that Abell2744-22741 is dominated by very young stellar populations or an AGN which is able to produce an intense radiation field. 
We see additional evidence of intense radiation field from the rest-frame UV (G140M) spectrum. 
We detect the [C~{\small III}], C~{\small III}] doublet emission. 
The total C~{\small III}] EW is $25\pm4$~\AA, greater than the majority of $z\sim7$ galaxies (median C~{\small III}] EW $\simeq8$~\AA; \citealt{Roberts-Borsani2024,Tang2026}).
We also identify Ly$\alpha$ emission line (EW $=17\pm2$~\AA) in the G140M spectrum. 
The Ly$\alpha$ peak flux is offset by $+227\pm92$~km~s$^{-1}$ from the line center, comparable to the typical Ly$\alpha$ velocity offset of $z>6$ galaxies \citep[e.g.,][]{Saxena2024,Tang2024b}. 
If the intense radiation field of Abell2744-22741 is able to enhance the ionization fraction of the surrounding IGM, it may boost the Ly$\alpha$ transmission of this system and even that of Abell2744-QSO1. 

Abell2744-25830 is much fainter, with a UV luminosity (M$_{\rm UV}=-17.0$) comparable to that of Abell2744-QSO1. 
This galaxy is even closer to Abell2744-QSO1, which is only $0.16$~pMpc away in the source plane. 
Using the rest-frame optical emission line detections (H$\beta$, [O~{\small III}]~$\lambda4959$, [O~{\small III}]~$\lambda5007$, H$\alpha$) of Abell2744-25830 (right panels of Figure~\ref{fig:z7_lae_spec}), we derive a systemic redshift of $z_{\rm sys}=7.0301$. 
The rest-frame optical emission lines of this system are extremely strong. 
We derive a very large [O~{\small III}]+H$\beta$ EW ($3518\pm217$~\AA), with a value among the upper $1\%$ of EWs observed at $z\sim7$ \citep{Endsley2024}. 
Such strong [O~{\small III}]+H$\beta$ emission is often linked to strong C~{\small IV} emission in cases where the metallicity is low \citep{Topping2025}. 
In the rest-frame UV spectrum of Abell2744-25830, we detect a C~{\small IV} doublet with an extremely large EW ($68\pm6$~\AA). 
This indicates a very hard radiation field, which may potentially create a large reionized bubble. 
We identify a strong Ly$\alpha$ emission for Abell2744-25830. 
Its Ly$\alpha$ peak flux is only offset by $+123\pm92$~km~s$^{-1}$ from the line center, smaller than the majority of the Ly$\alpha$ emitters at $z>7$.
The Ly$\alpha$ EW ($166\pm9$~\AA) is atypically large among the $z\sim7$ population \citep[e.g.,][]{Napolitano2024,Tang2024c,Jones2025}, as expected if the Ly$\alpha$ production is boosted by the hard radiation field and/or the transmission is enhanced in large ionized bubble. 

To summarize, both the two newly identified $z\simeq7.04$ galaxies nearby Abell2744-QSO1 appear to have intense radiation fields, especially Abell2744-25830 which is closer to Abell2744-QSO1. 
These two systems may be part of an overdensity of galaxies that is contributing to the reionization of the IGM surrounding Abell2744-QSO1, aiding the Ly$\alpha$ to escape when the Universe was partially neutral. 
We will describe the SPURS Ly$\alpha$ measurement of Abell2744-QSO1 in Section~\ref{sec:QSO1_uv_spec} and discuss the Ly$\alpha$ profile in Section~\ref{sec:lya_model}.

\section{Deep Rest-Frame UV Spectra of LRDs} \label{sec:lrd_uv_spec}

\subsection{Abell2744-QSO1} \label{sec:QSO1_uv_spec}


\begin{deluxetable}{cccc}
\tablecaption{Rest-frame UV emission line flux ($\times10^{-20}$~erg~s$^{-1}$~cm$^{-2}$), EW (\AA), and FWHM (km~s$^{-1}$) of Abell2744-QSO1 measured from SPURS spectra.}
\tablehead{
Line & Flux & EW & FWHM
}
\startdata
narrow Ly$\alpha$ & $85.2\pm8.1$ & $23.8\pm2.3$ & $333\pm48$ \\
broad Ly$\alpha$ & $319.0\pm15.6$ & $88.9\pm4.3$ & $1498\pm145$ \\
N~{\scriptsize V}~$\lambda1239$ & $<15.5$ & $<4.3$ & - \\
N~{\scriptsize V}~$\lambda1243$ & $<15.7$ & $<4.4$ & - \\
O~{\scriptsize I}~$\lambda1302$ & $17.6\pm5.4$ & $3.6\pm1.1$ & $348\pm95$ \\
{[}N~{\scriptsize IV}]~$\lambda1483$ & $<13.9$ & $<3.2$ & - \\
N~{\scriptsize IV}]~$\lambda1486$ & $<13.2$ & $<3.0$ & - \\
C~{\scriptsize IV}~$\lambda1549^{\rm b}$ & $25.4\pm6.0$ & $5.7\pm1.3$ & - \\
He~{\scriptsize II}~$\lambda1640^{\rm a}$ & $<21.8$ & $<5.4$ & - \\
O~{\scriptsize III}]~$\lambda1661$ & $<9.0$ & $<2.2$ & - \\
O~{\scriptsize III}]~$\lambda1666$ & $<8.4$ & $<2.1$ & - \\
Fe~{\scriptsize II}~$\lambda1786$ & $23.7\pm5.6$ & $5.7\pm1.4$ & - \\
{[}C~{\scriptsize III}]~$\lambda1907$ & $<10.3$ & $<2.9$ & - \\
C~{\scriptsize III}]~$\lambda1909$ & $<10.7$ & $<3.0$ & - \\
{[}Ne~{\scriptsize IV}]~$\lambda2422$ & $<25.4$ & $<8.8$ & - \\
{[}Ne~{\scriptsize IV}]~$\lambda2424$ & $<25.4$ & $<8.8$ & - \\
Mg~{\scriptsize II}~$\lambda2800^{\rm a}$ & $<42.5$ & $<13.0$ & - \\
{[}Ne~{\scriptsize V}]~$\lambda3427$ & $<19.0$ & $<11.9$ & - \\
\enddata
\tablecomments{Fluxes are not corrected for gravitational magnification. We show $3\sigma$ upper limits for non-detections. \\ $^{\rm a}$: Upper limits ($3\sigma$) of line flux and EW of broad permitted, He~{\scriptsize II}, and Mg~{\scriptsize II} emission, assuming FWHM $\simeq2600$~km~s$^{-1}$. \\ $^{\rm b}$: Total line flux and EW of the C~{\scriptsize IV}~$\lambda\lambda1548,1551$ doublet.}
\label{tab:QSO1_uv_line}
\end{deluxetable}

We now discuss the features in the SPURS rest-frame UV spectrum of Abell2744-QSO1 (Figure~\ref{fig:QSO1_uv_spec}). 
We consider both the G140M and part of the G235M data, probing rest-frame wavelengths of $1200-2050$~\AA\ and $2030-3500$~\AA, respectively. 
We visually search the spectrum for emission lines using the systemic redshift ($z_{\rm sys}=7.0364$, Section~\ref{sec:QSO1_info}). 
An extremely broad Ly$\alpha$ emission line profile is detected with its flux peaking at an observed wavelength of $9778$~\AA\ (S/N $=23$). 
We identify iron emission multiplets, Fe~{\small II} near rest-frame $1786$~\AA\ (S/N $=4$). 
We detect the C~{\small IV} emission line (S/N $=4$), although we do not cleanly resolve the doublet. 
We additionally report an emission line at observed wavelength of $10460$~\AA\ (S/N $=3$), close to the expected position of O~{\small I}~$\lambda1302$. 
We report the rest-frame UV emission line flux, EW, and FWHM measurements in Table~\ref{tab:QSO1_uv_line}.

\subsubsection{Lya Emission} \label{sec:QSO1_Lya}

We measure a total Ly$\alpha$ EW of $113\pm5$~\AA, consistent with that measured from the R100 prism spectrum \citep{Furtak2024,Ji2025}. 
With the SPURS R1000 spectrum, we characterize the Ly$\alpha$ velocity profile (i.e., peak velocity offset, line width).
We find that the line profile is asymmetric, with a sharp cutoff near the line center, as expected if the blue Ly$\alpha$ emission is strongly attenuated by the partially neutral IGM at $z\simeq7$ \citep[e.g.,][]{Dijkstra2007,Saxena2024,Tang2024c}. 


\begin{figure*}
\includegraphics[width=\linewidth]{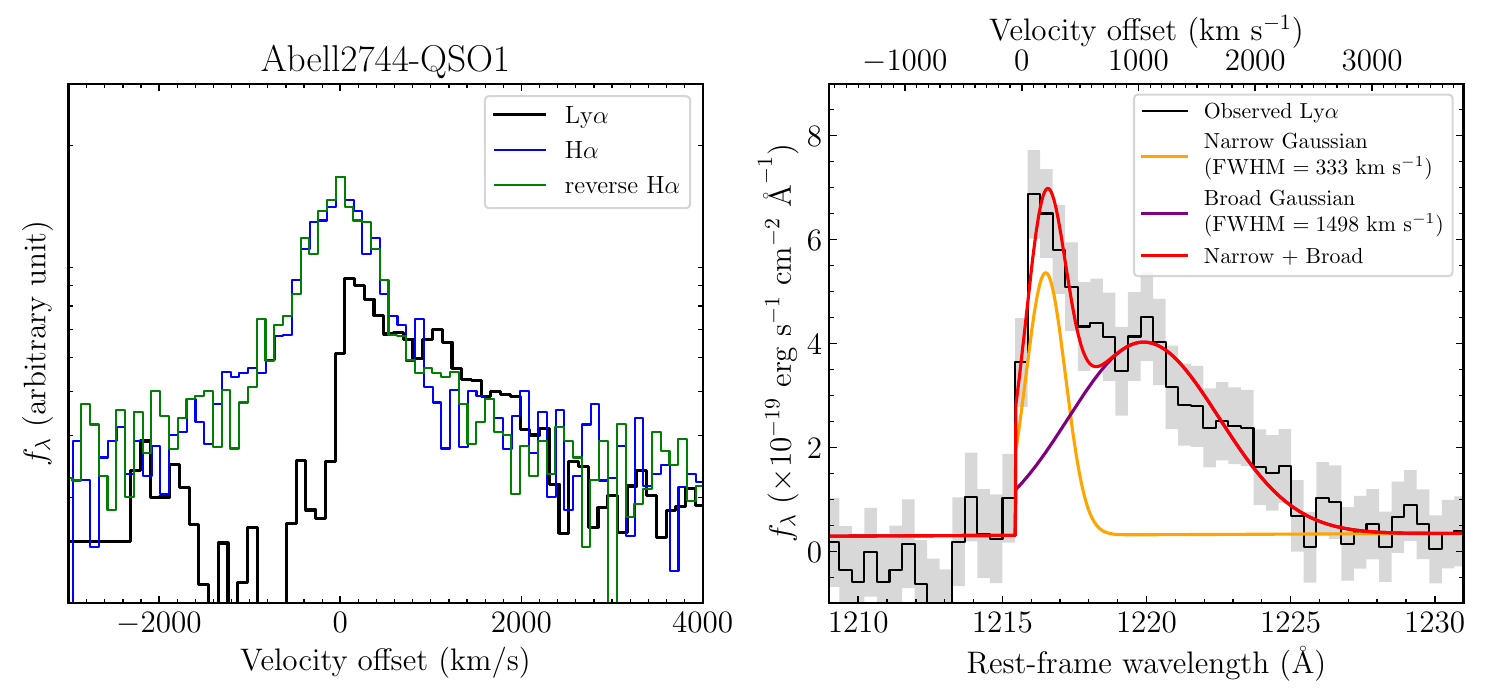}
\caption{Ly$\alpha$ emission line profile of Abell2744-QSO1. In the left panel, we show Ly$\alpha$ (black) in the velocity space, overplotting the H$\alpha$ profile (blue) and its reversed velocity profile (green). Both the Ly$\alpha$ and H$\alpha$ lines are normalized by the flux densities at velocities $2000-2500$~km~s$^{-1}$ redshifted from the line center. To better compare the wing profiles of Ly$\alpha$ and H$\alpha$ emission, the flux density (y-axis) is shown in logarithmic scale. In the right panel, we show the empirical two truncated Gaussian fitting (red) to the observed Ly$\alpha$ profile (black). The Ly$\alpha$ can be naively fitted by a narrow (FWHM $=333$~km~s$^{-1}$; orange) and a broad (FWHM $=1498$~km~s$^{-1}$; purple) component. We will discuss the Ly$\alpha$ profile in more detail in Section~\ref{sec:lya_model}.}
\label{fig:QSO1_lya}
\end{figure*}

The red side of the Ly$\alpha$ profile is very broad. 
In particular, we find a red tail of emission extending to $\simeq2500$~km~s$^{-1}$ from the systemic redshift. 
This is near-identical to the width of the red-side wing of the broad H$\alpha$ line which we illustrate in the left panel of Figure~\ref{fig:QSO1_lya}. 
The broad component of Ly$\alpha$ appears superimposed on a narrower component peaking closer to line center. 
To obtain a first-order empirical description of the Ly$\alpha$ emission, we begin by fitting the line profile with two truncated Gaussian functions \citep[e.g.,][]{Endsley2022a}, as shown in the right panel of Figure~\ref{fig:QSO1_lya}. 
We derive a FWHM of $333\pm48$~km~s$^{-1}$ for the narrow component, comparable to the instrument resolution ($\simeq300$~km~s$^{-1}$) and centered at $+258\pm92$~km~s$^{-1}$.
The broad component has a FWHM of $1498\pm145$~km~s$^{-1}$, with its peak flux further redshifted (velocity offset $=+1015\pm92$~km~s$^{-1}$). 

\begin{figure}
\includegraphics[width=\linewidth]{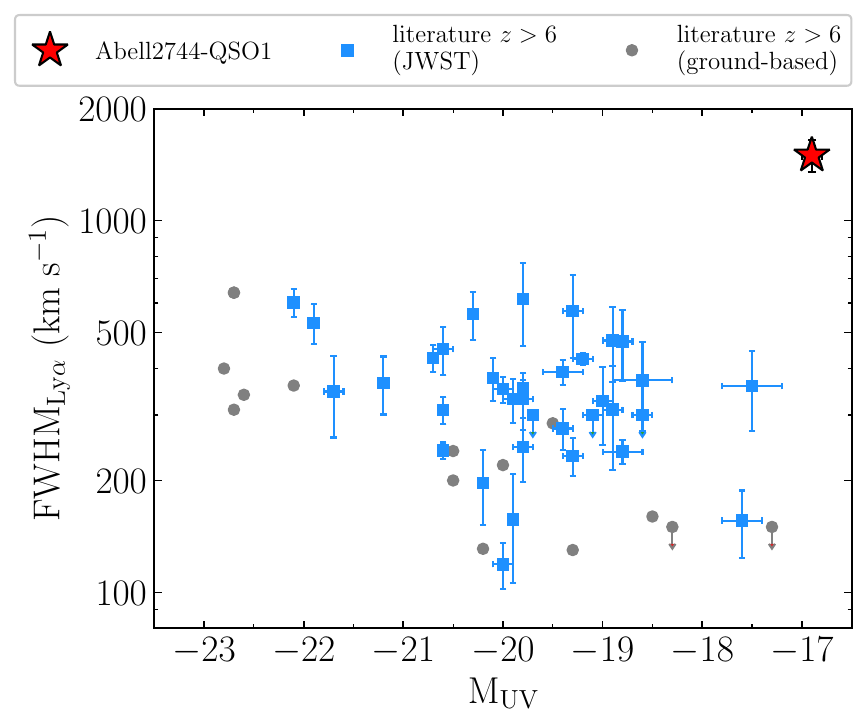}
\caption{Ly$\alpha$ FWHM versus M$_{\rm UV}$ of $z>6$ galaxies. The red star represents Abell2744-QSO1. As a comparison, we overplot data from literature. The cyan squares show systems with Ly$\alpha$ FWHM measured from JWST/NIRSpec R1000 or R2700 grating spectra \citep{Bunker2023,Tang2024c,Chen2025,Whitler2025,Witstok2025}. The grey circles show galaxies observed from ground-based facilities (see \citealt{Endsley2022b} and references therein). The Ly$\alpha$ line of Abell2744-QSO1 is much broader than that of other galaxies at similar M$_{\rm UV}$.}
\label{fig:lya_fwhm}
\end{figure}

Prior to the SPURS observations, one possibility was that Ly$\alpha$ in Abell2744-QSO1 was entirely associated with ionized gas in the host galaxy. 
In this case, we would have expected a narrow line profile, similar to that seen in other faint star forming galaxies at high redshift. 
In Figure~\ref{fig:lya_fwhm}, we show the Ly$\alpha$ FWHM as a function of absolute UV magnitude for a compilation of high redshift galaxies. 
For extremely low luminosity hosts like Abell2744-QSO1, we expect the Ly$\alpha$ FWHM to be at most a few hundred km~s$^{-1}$. 
It is clear that Abell2744-QSO1 has a much broader Ly$\alpha$ line than expected, with FWHM $5-10\times$ larger than that of galaxies with similar UV luminosities at $z>6$. 
This analysis suggests that there is likely a non-host component associated with the Ly$\alpha$ emission. 
Or if Ly$\alpha$ is from the host, it must be very different than typical star forming systems. 
Perhaps a more natural interpretation is that the broad Ly$\alpha$ is linked to whatever mechanism is producing the broad Balmer lines.
We will come back to discuss the Ly$\alpha$ profile of Abell2744-QSO1 in more detail in Section~\ref{sec:lya_model}. 

We can estimate the escape fraction ($f_{\rm esc,Ly\alpha}$) of both the narrow and broad components of Ly$\alpha$ using the (dust-corrected) H$\alpha$ lines to predict the intrinsic Ly$\alpha$ luminosity ($L^{\rm int}_{\rm Ly\alpha}=8.7\times L_{\rm H\alpha}$; e.g., \citealt{Hu1998,Hayes2015,Henry2015}). 
As we will note below, such estimates face a range of uncertainties, but they nevertheless will help guide possible explanations for the origin of the line emission.
We first focus on the escape fraction of the narrow Ly$\alpha$ line, which we may assume is the host galaxy component.
We measure a narrow Ly$\alpha$ line flux of $8.5\pm0.8\times10^{-19}$~erg~s$^{-1}$~cm$^{-2}$ using the decomposition described above.
Assuming the Small Magellanic Cloud (SMC) extinction law \citep{Gordon2003} and case B recombination (intrinsic H$\alpha$/H$\beta$ ratio $=2.87$; \citealt{Osterbrock2006}), the observed narrow H$\alpha$/H$\beta$ ratio ($3.9\pm1.7$; Table~\ref{tab:QSO1_opt_line}) implies an attenuation of $A_{\rm H\alpha,n}=0.7\pm0.3$~mag to the narrow H$\alpha$ line.
We then derive an escape fraction of $0.086\pm0.026$ for the narrow Ly$\alpha$ line. 
If we instead assume no attenuation, our results suggest an escape fraction of $0.16\pm0.05$. 
In both cases, this calculation indicates the bulk of the narrow Ly$\alpha$ line photons are not entering the NIRSpec microshutter. 
This is common in $z>7$ galaxies, owing in part to IGM attenuation.

We can follow the same approach for the broad component of Ly$\alpha$. 
Our decomposition suggests a broad line flux of $3.2\pm0.2\times10^{-18}$~erg~s$^{-1}$~cm$^{-2}$. 
To estimate the escape fraction of broad Ly$\alpha$ photons, we require a measurement of the broad Balmer line luminosity, which in turn requires constraints on the attenuation facing the broad Balmer lines. 
This is non-trivial owing to the possibility that the Balmer decrement is impacted by scattering and collisional-excitation \citep[e.g.,][]{deGraaff2025c,DEugenio2025b,Nikopoulos2025,Chang2026}. 
If we assume the SMC law and an intrinsic broad H$\alpha$/H$\beta$ ratio of $3.06$ \citep{Dong2008} that is adopted in previous studies \citep{DEugenio2025a,Ji2025}, we find an attenuation of $A_{\rm H\alpha,b}=1.8\pm0.6$~mag to the broad H$\alpha$ line using the broad H$\alpha$/H$\beta$ ratio ($7.0\pm2.3$; Table~\ref{tab:QSO1_opt_line}). 
However, as we noted above, it is possible that the Balmer decrement does not reflect dust attenuation, so we also consider the case where the broad lines do not face attenuation. 
To predict the intrinsic Ly$\alpha$ luminosity, we furthermore assume case B recombination, which introduces more uncertainty. 
In the case of modest dust attenuation, we find the broad line region has a Ly$\alpha$ escape fraction of $0.007\pm0.002$. 
If the broad line attenuation is zero, the inferred broad Ly$\alpha$ escape fraction is $0.034\pm0.009$. 
In both cases, we find that the observed broad Ly$\alpha$ photons are likely a small fraction of the intrinsic line output. 

Finally, we consider the possibility that the entire Ly$\alpha$ profile (narrow and broad components) has its origin in the same gas that produces the narrow Balmer line. 
This may be expected if the origin of the Ly$\alpha$ emission is not related to the broad Balmer lines, but instead is powered by ionizing sources in the host galaxy. 
Here we estimate a Ly$\alpha$ escape fraction of $0.75\pm0.21$ assuming no dust attenuation and $0.41\pm0.12$ in the case of modest attenuation implied by the narrow line Balmer decrement.


\subsubsection{High Ionization Lines and Broad Emission Lines} \label{sec:QSO1_high_ion}

The SPURS dataset provides our most stringent constraints on high ionization lines in Abell2744-QSO1. 
The detection of C~{\small IV} emission suggests the presence of a hard radiation field, supplying ionizing photons with energies $>48$~eV.
We measure a total C~{\small IV} flux of $2.5\pm0.6\times10^{-19}$~erg~s$^{-1}$~cm$^{-2}$, indicating an EW of $5.7\pm1.3$~\AA. 
Such narrow line C~{\small IV} EWs are commonly seen in metal poor star forming galaxies \citep[e.g.,][]{Berg2019,Senchyna2019,Topping2025}. 

Given the presence of broad Balmer and Ly$\alpha$ emission lines in the spectrum of Abell2744-QSO1, we may naively expect the permitted C~{\small IV} emission to show a broad component as well. 
At the S/N ($=4$) of the detected emission, we cannot unambiguously resolve the doublet. 
We measure a FWHM of $906$~km~s$^{-1}$ for the C~{\small IV} emission feature, which is $\simeq3\times$ narrower than the widths of broad Balmer lines (FWHM$_{\rm H\alpha,broad}=2653\pm345$~km~s$^{-1}$). 
On the other hand, the observed C~{\small IV} profile is fully consistent with a blended narrow C~{\small IV}~$\lambda1548,1551$ doublet whose individual components have FWHMs comparable to the instrument resolution. 
Under the assumption that the detected C~{\small IV} is associated with the narrow line spectrum, we place an upper limit on the broad C~{\small IV} emission line flux of Abell2744-QSO1 assuming a FWHM$_{\rm broad}$ of $2600$~km~s$^{-1}$ and integrating the error spectrum in quadrature with a spectral window spanning $2\times {\rm FWHM}_{\rm broad}$. 
This implies a $3\sigma$ limiting flux of $<2.3\times10^{-19}$~erg~s$^{-1}$~cm$^{-2}$, indicating a broad C~{\small IV}/H$\beta$ flux ratio of $<0.15$. 
This is well below the typical C~{\small IV}/H$\beta$ flux ratio of type I AGN ($\sim3$; e.g., \citealt{Francis1991,Brotherton2001,VandenBerk2001}). 

We also constrain the permitted He~{\small II}~$\lambda1640$ and Mg~{\small II} lines, both of which could also plausibly be produced in the same line emitting region as the broad Balmer lines.
However, we do not detect either broad feature in the SPURS spectrum. 
Following the same approach as we followed for constraining the broad C~{\small IV} emission, we place a $3\sigma$ limiting flux of $<2.2\times10^{-19}$~erg~s$^{-1}$~cm$^{-2}$ to the broad He~{\small II} emission. 
This indicates a broad He~{\small II}/H$\beta$ flux ratio of $<0.14$, below that which is typical of type I AGN ($\sim0.5$; e.g., \citealt{Francis1991,Brotherton2001}). 
For the broad Mg~{\small II} emission, we derive a $3\sigma$ limiting flux of $<4.3\times10^{-19}$~erg~s$^{-1}$~cm$^{-2}$. 
The corresponding broad Mg~{\small II}/H$\beta$ flux ratio is $<0.28$, more than 5 times lower than that seen in typical type I AGN ($\sim1.3-1.7$; e.g., \citealt{Francis1991,Brotherton2001,VandenBerk2001}). 
While Abell2744-QSO1 appears able to produce an emergent broad Ly$\alpha$ line, the other permitted broad lines in the rest-frame UV are not seen. 

As the presence of narrow C~{\small IV} emission reveals a hard radiation field Abell2744-QSO1, we may expect to detect line emission from other highly-ionized species. 
The G140M spectrum covers N~{\small IV}], O~{\small III}], and C~{\small III}], all emission lines that are seen in metal poor galaxy spectra, particularly those also showing C~{\small IV} emission. 
However none of these lines are detected. 
For the above forbidden lines, we constrain the line flux of each individual component assuming a narrow line (i.e., FWHM is comparable to the instrument resolution, $300$~km~s$^{-1}$). 
The limits are listed in Table~\ref{tab:QSO1_uv_line}. 
In particular, we place a $3\sigma$ limiting flux of $<1.0\times10^{-19}$~erg~s$^{-1}$~cm$^{-2}$ for each component of the C~{\small III}] doublet, indicating a line EW that is below $3$~\AA. 
Comparing with the C~{\small IV} detection, the C~{\small III}]/C~{\small IV} flux ratio is below $0.8$ at $3\sigma$, consistent with photoionization models driven by either massive stars or AGN \citep[e.g.,][]{Gutkin2016,Feltre2016,Mignoli2019,Plat2019}. 
We additionally constrain the line flux of the narrow He~{\small II} emission, placing a $3\sigma$ limit of $<9.2\times10^{-20}$~erg~s$^{-1}$~cm$^{-2}$.
This indicates a C~{\small IV}/He~{\small II} flux ratio of $>2.8$ at $3\sigma$, consistent with photoionization models driven by massive stars \citep[e.g.,][]{Gutkin2016,Feltre2016}.

Finally, we also do not identify higher ionization potential lines (N~{\small V}, [Ne~{\small IV}], [Ne~{\small V}]) for Abell2744-QSO1. 
With the G140M spectrum, we place a $3\sigma$ limiting EW of $<4$~\AA\ for each of the individual components of the N~{\small V} doublet.
At the redshift of Abell2744-QSO1, [Ne~{\small IV}]~$\lambda\lambda2422,2424$ and [Ne~{\small V}]~$\lambda3427$ emission lines will be shifted to the G235M spectrum. 
Our results suggest that each individual component of [Ne~{\small IV}] ([Ne~{\small V}]) has an EW below $9$~\AA\ ($12$~\AA) at $3\sigma$. 
The EW limits are similar to those measured from the small number of LRDs with the deepest rest-frame UV grating coverage \citep[e.g.,][]{Tang2025}. 
Given that there are not clear signatures of AGN photoionization in Abell2744-QSO1, we suggest that the narrow C~{\small IV} emission is plausibly associated with a low metallicity massive stellar population in the host galaxy, as would be expected based on the low metallicities implied by the weak [O~{\small III}] emission. 
However, we cannot rule out an association with the LRD. 

\subsubsection{O~{\scriptsize I} and Fe~{\scriptsize II} Emission} \label{sec:QSO1_OI_FeII}

O~{\small I} and Fe~{\small II} emission lines are often present in the spectra of AGN \citep[e.g.,][]{Grandi1980,Wills1985,Rodriguez-Ardila2002,Juodzbalis2024,Tripodi2025,Torralba2026b}. 
However, these lines have not been reported for Abell2744-QSO1 in previous papers.
Because of the similarity of the Ly$\beta$ resonance wavelength and that of the 3d$^3$ D$^0$ excited state of the O~{\small I} atom, the strength of O~{\small I} can be enhanced significantly by Ly$\beta$ fluorescence \citep{Kwan1981}. 
We detect O~{\small I}~$\lambda1302$ emission in the G140M spectrum of Abell2744-QSO1 with a line EW of $3.6\pm1.1$~\AA, which may be boosted by Ly$\beta$ fluorescence in extremely dense gas. 
Similarly, the Fe~{\small II} emission can be enhanced by Ly$\alpha$ fluorescence. 
We derive an Fe~{\small II}~$\lambda1786$ EW of $5.7\pm1.4$~\AA, stronger than that of typical type I AGN \citep[e.g.,][]{VandenBerk2001}. 
These detections suggest that Ly$\alpha$ and Ly$\beta$ must be highly optically thick, perhaps consistent with the low Ly$\alpha$ escape fractions described above. 

\subsubsection{UV Continuum Slope and Absorption Lines} \label{sec:QSO1_cont_abs}

In addition to emission lines, the rest-frame UV continuum of Abell2744-QSO1 is marginally detected (median S/N $=2$ per resolution element) in the SPURS spectrum, allowing us to quantify the UV slope and constrain the interstellar as well as stellar absorption features. 
We measure a relatively blue UV continuum slope of $\beta_{\rm UV}=-1.5\pm0.3$, consistent with that inferred from NIRSpec prism spectrum ($\beta_{\rm UV}=-1.5$) and NIRCam \citep{Rieke2023} broadband photometry ($\beta_{\rm UV}=-1.6$; \citealt{Furtak2023a,Furtak2024}). 

Depending on the nature of the sources dominating the UV continuum, we may expect to see absorption lines from interstellar gas and massive stars. 
We first visually search the G140M spectrum for rest-frame UV interstellar absorption lines (Si~{\small II}~$\lambda1260$, O~{\small I}~$\lambda1302$, C~{\small II}~$\lambda1334$, Si~{\small IV}~$\lambda\lambda1393,1402$, Si~{\small II}~$\lambda1526$, C~{\small IV}~$\lambda1549$, Al~{\small II}~$\lambda1670$), but none of them are detected. 
Here we note that because O~{\small I} and C~{\small IV} emission is present, the underlying absorption components might be filled in by emission. 
We constrain the strength of UV interstellar absorption features, placing a $3\sigma$ upper limit on EW of $\simeq-2.7$~\AA\ for each individual line.
This is consistent with the interstellar absorption line EW measured from composite spectra of galaxies at $z\sim3-7$ ($-2$ to $-1$~\AA; e.g., \citealt{Shapley2003,Jones2012,Steidel2016,Pahl2020,Glazer2025}). 
Based on these results, our spectrum suggests that interstellar lines are likely somewhat weak in Abell2744-QSO1, which is perhaps not surprising given the low gas covering fractions which are typical at $z>7$ \citep{Glazer2025} and the extremely low metallicity of the host.

Stellar absorption features would provide the most reliable signature that the host galaxy dominates the UV continuum. 
We search for wind and photospheric absorption features (N~{\small V}, 1302 index, 1370 index, 1425 index, C~{\small IV}, 1978 index; \citealt{Rix2004,Steidel2016,Vidal-Garcia2017}). 
Neither is clearly detected in the G140M spectrum. 
However, we note that the limits on photospheric absorption are not sufficiently constraining. 
Future observations with deeper spectroscopy are required to put more robust constraints on the rest-frame UV absorption features of Abell2744-QSO1. 

\subsection{UNCOVER-2476} \label{sec:2476_uv_spec}

We characterize the rest-frame UV spectrum of UNCOVER-2476, which is partially covered by G140M (rest-frame $1930-3500$~\AA) in the SPURS observations. 
Using the systemic redshift ($z_{\rm sys}=4.0197$, Section~\ref{sec:2476_info}), we visually search the G140M spectrum (Figure~\ref{fig:2476_uv_spec}) for emission lines. 
The most prominent rest-frame UV line detection is at $14039$~\AA\ (S/N $=8$) with a companion at $14074$~\AA\ (S/N $=3$), close to the expected wavelengths of Mg~{\small II}~$\lambda\lambda2796,2803$ doublet.
We identify two iron emission lines [Fe~{\small IV}]~$\lambda2829$ (S/N $=4$) and [Fe~{\small IV}]~$\lambda2835$ (S/N $=6$). 
We additionally detect an emission feature at $12169$~\AA\ (S/N $=4$), consistent with the expected position of [Ne~{\small IV}]~$\lambda\lambda2422,2424$. 
Helium emission lines at rest-frame near-UV are also detected, including He~{\small II}~$\lambda2733$ (S/N $=4$), He~{\small I}~$\lambda3188$ (S/N $=4$), and tentative He~{\small II}~$\lambda3203$ (S/N $=2$). 
We report the rest-frame UV emission line measurements in Table~\ref{tab:2476_uv_line}. 
Along with emission lines, the rest-frame UV continuum is clearly present in the G140M spectrum with a median S/N of $6$ per resolution element, allowing us to characterize absorption lines. 
In particular, we find an Fe~{\small II}~$\lambda2586$ absorption feature (S/N $=4$), but will also discuss other possible absorption features throughout the rest-frame UV. 


\begin{deluxetable}{cccc}
\tablecaption{Rest-frame UV emission line flux ($\times10^{-20}$~erg~s$^{-1}$~cm$^{-2}$), EW (\AA), and FWHM (km~s$^{-1}$) of UNCOVER-2476 measured from SPURS spectra.}
\tablehead{
Line & Flux & EW & FWHM
}
\startdata
{[}Ne~{\scriptsize IV}]$^{\rm a}$ & $17.9\pm5.1$ & $1.3\pm0.3$ & - \\
He~{\scriptsize II}~$\lambda2733$ & $18.2\pm4.4$ & $1.7\pm0.4$ & $351\pm85$ \\
Mg~{\scriptsize II}~$\lambda2796$ & $28.6\pm3.8$ & $2.8\pm0.4$ & $216\pm75$ \\
Mg~{\scriptsize II}~$\lambda2803$ & $11.5\pm3.6$ & $1.2\pm0.4$ & $135\pm75$ \\
{[}Fe~{\scriptsize IV}]~$\lambda2829$ & $12.6\pm3.3$ & $1.3\pm0.3$ & $227\pm74$ \\
{[}Fe~{\scriptsize IV}]~$\lambda2835$ & $21.8\pm3.7$ & $2.3\pm0.4$ & $210\pm74$ \\
He~{\scriptsize I}~$\lambda3188$ & $17.3\pm3.5$ & $2.2\pm0.5$ & $182\pm66$ \\
He~{\scriptsize II}~$\lambda3203$ & $5.8\pm2.4$ & $0.7\pm0.3$ & $118\pm65$ \\
{[}Ne~{\scriptsize V}]~$\lambda3427$ & $<12.0$ & $<1.7$ & - \\
\enddata
\tablecomments{Fluxes are not corrected for gravitational magnification. We show $3\sigma$ upper limits for non-detections. \\ $^{\rm a}$: Total line flux and EW of the [Ne~{\scriptsize IV}]~$\lambda\lambda2422,2424$ doublet.}
\label{tab:2476_uv_line}
\end{deluxetable}

\subsubsection{High Ionization Lines} \label{sec:2476_high_ion}

The detection of [Ne~{\small IV}]~$\lambda\lambda2422,2424$ emission either points to the presence of a hard radiation field with photons having energies $>64$~eV or fast-radiative shocks. 
As massive stars do not emit many photons at these energies, the former case may require the escape of hard photons from an AGN \citep[e.g.,][]{Feltre2016,Mignoli2019,Terao2022}. 
On the other hand, shocks may be expected from dense outflowing and turbulent gas. 
We measure a total [Ne~{\small IV}] doublet flux of $1.8\pm0.5\times10^{-19}$~erg~s$^{-1}$~cm$^{-2}$, corresponding to an EW of $1.3\pm0.3$~\AA. 
The [Ne~{\small IV}] EW of UNCOVER-2476 is lower than that of other [Ne~{\small IV}] emission lines identified from JWST/NIRSpec observations (EW $\simeq10-20$~\AA; \citealt{Maiolino2024a,Tang2025}), but such weak lines are common among type I AGN \citep[e.g.,][]{Francis1991,VandenBerk2001,Mignoli2019}. 
We do not detect emission from more highly ionized neon species ([Ne~{\small V}]). 
We place a $3\sigma$ limiting flux of $<1.2\times10^{-19}$~erg~s$^{-1}$~cm$^{-2}$ on [Ne~{\small V}]~$\lambda3427$ emission, indicating an EW limit $<1.7$~\AA. 

We also find high ionization emission from [Fe~{\small IV}] and He~{\small II} in the rest-frame near-UV. 
Both components of the [Fe~{\small IV}]~$\lambda\lambda2829,2835$ doublet are narrow, with FWHM ($210-227$~km~s$^{-1}$) comparable to the instrument resolution within $1\sigma$ uncertainty. 
Such narrow [Fe~{\small IV}] emission lines are occasionally found in type II AGN \citep[e.g.,][]{Rose2011}. 
The He~{\small II}~$\lambda2733$ emission line is associated with the $n=6$ to $n=3$ transition of He$^+$. 
We derive an EW of $1.7\pm0.4$~\AA\ for He~{\small II}~$\lambda2733$, consistent with the typical EW seen in type II AGN \citep[e.g.,][]{Zakamska2003}. 

\subsubsection{Mg~{\scriptsize II} Emission} \label{sec:2476_MgII}

The SPURS G140M spectrum reveals narrow Mg~{\small II} emission lines (FWHM $=135$~km~s$^{-1}$ for Mg~{\small II}~$\lambda2796$ and $216$~km~s$^{-1}$ for Mg~{\small II}~$\lambda2803$). 
We derive an EW of $2.8\pm0.4$~\AA\ for Mg~{\small II}~$\lambda2796$ emission and $1.2\pm0.4$~\AA\ for Mg~{\small II}~$\lambda2803$ emission, comparable to the Mg~{\small II} EWs measured in the spectra of low-metallicity star-forming galaxies \citep[e.g.,][]{Guseva2013,Izotov2016,Izotov2018,Henry2018}. 
On the other hand, we do not detect broad Mg~{\small II} emission. 
Assuming a line width that is comparable to the broad Balmer emission lines (FWHM $=2000$~km~s$^{-1}$; Table~\ref{tab:2476_opt_nir_line}), we constrain a $3\sigma$ limiting flux of $<3.0\times10^{-19}$~erg~s$^{-1}$~cm$^{-2}$ for the broad Mg~{\small II}. 
This indicates a broad Mg~{\small II}/H$\beta$ flux ratio of $<0.14$, well below the ratios seen in typical type I AGN ($\sim1.3-1.7$; e.g., \citealt{Francis1991,Brotherton2001,VandenBerk2001}). 

The Mg~{\small II} emission also provides insight into the gas conditions of UNCOVER-2476. 
The Mg~{\small II} line fluxes indicate a doublet ratio of $f_{{\rm MgII}\lambda2796}/f_{{\rm MgII}\lambda2803}=2.5\pm0.8$.
This is close to the intrinsic Mg~{\small II} doublet ratio ($2$) when collisions dominate the Mg$^+$ excitation, as expected from optically thin Mg~{\small II} gas \citep[e.g.,][]{Chisholm2020,Chang2024}. 
We note that the Mg~{\small II} emission can be pumped, but the Mg~{\small II} pumping itself should not alter the doublet ratio. 
We also find that the Mg~{\small II} velocity offset is small. 
The flux peak of Mg~{\small II}~$\lambda2796$ (Mg~{\small II}~$\lambda2803$) is just $+32\pm75$~km~s$^{-1}$ ($+96\pm75$~km~s$^{-1}$) offset from the line center, as expected if Mg$^+$ photons have experienced minimal resonant scattering in optically thin gas. 
The ionized gas in the narrow-line emitting region (plausibly ionized gas in the host galaxy) appears to be minimally covered by dense neutral gas.

\subsubsection{UV Continuum Slope and Absorption Lines} \label{sec:2476_cont_abs}

The rest-frame UV continuum detection allows us to measure the UV slope.
While it is common to derive a UV slope by fitting the continuum spectrum at rest-frame $1250-2600$~\AA\ \citep[e.g.,][]{Calzetti1994}, our G140M spectrum only covers rest-frame wavelength $>1930$~\AA, so we derive the UV slope by fitting the continuum spectrum at rest-frame $1930-2600$~\AA. 
Nevertheless, we still find a UV slope ($\beta_{\rm UV}=-1.82\pm0.03$) that is consistent with the photometric measurement ($\beta_{\rm UV}=-1.76\pm0.01$; \citealt{Labbe2025}). 
This indicates that UNCOVER-2476 has a bluer UV continuum relative to much of the LRD population (median $\beta_{\rm UV}=-1.43$; \citealt{deGraaff2025c}). 
It is plausible this suggests a more dominant host galaxy contribution to the UV, consistent with the weaker Balmer break.

The high S/N of UV continuum (median S/N $=6$ per resolution element) also enables us to constrain the photospheric absorption lines from OB stars, potentially providing insight into the contribution of the host galaxy. 
One of the most commonly used indices at rest-frame near-UV wavelengths is the 1978 index, which is dominated by iron photospheric lines. 
We calculate the EW of the 1978 index using the methods in \citet{Rix2004}, getting an EW of $-3.6\pm1.9$~\AA. 
This indicates a tentative detection (S/N $=2$) of the 1978 index absorption feature, broadly consistent with the EW that is expected from low metallicity stars ($Z=0.07\ Z_{\odot}$, see Section~\ref{sec:2476_opt_nir_spec}; \citealt{Rix2004}). 
We also detect the Fe~{\small II}~$\lambda2586$ absorption line in the continuum spectrum, with Fe~{\small II}~$\lambda2586$ EW of $-3.2\pm0.8$~\AA. 
This is less prominent than the absorption line EWs of quasars ($\simeq-10$~\AA\ and even stronger; e.g., \citealt{Rodriguez-Hidalgo2011,Rafiee2016,DEugenio2025b}) but comparable to those that are commonly seen in star-forming galaxies ($\sim-3$ to $-1$~\AA; e.g., \citealt{Quider2009,Rubin2010,Finley2017}).


\begin{figure}
\includegraphics[width=\linewidth]{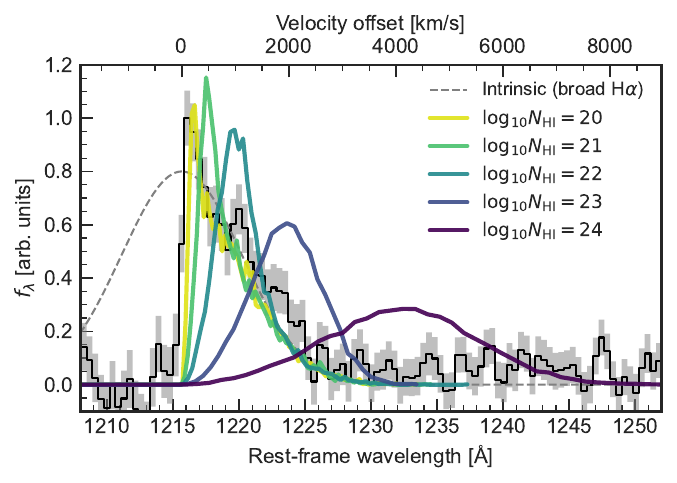}
\caption{Ly$\alpha$ spectra after resonant scattering through a static uniform shell with a range of H~{\scriptsize I} column densities (colored lines), assuming the intrinsic line profile is similar to the broad component of H$\alpha$ (FWHM $\approx2600$~km~s$^{-1}$, gray dashed line). We truncate the emergent profiles blueward of systemic velocity, as expected from scattering in the IGM at $z>6$. We also show the observed Ly$\alpha$ spectrum for comparison (black line). Due to the high cross-section for scattering in the damping wings, gas with $N_{\rm HI} > 10^{21}$~cm$^{-2}$ will be optically thick to photons far from Ly$\alpha$ line center and thus significantly redshift the emerging line shape. The noisiness of the model lines is due to numerical effects.}
\label{fig:Lya_tau}
\end{figure}

\section{Characterizing L\lowercase{y}$\alpha$ in Abell2744-QSO1} \label{sec:lya_model}

With our G140M spectrum we have characterized the Ly$\alpha$ emission line profile in Abell2744-QSO1, finding it has both a narrow component (FWHM $\sim300$~km~s$^{-1}$), centered at $\sim260$~km~s$^{-1}$ from line center, and an extremely broad component (FWHM $\sim1500$~km~s$^{-1}$), offset by $\sim1000$~km~s$^{-1}$ from line center (Section~\ref{sec:QSO1_uv_spec}), and a sharp cut-off blueward of line center.
As a resonant line, the velocity profile of Ly$\alpha$ is extremely sensitive to the properties of the gas through which the photons scatter. 
In particular, Ly$\alpha$ profiles are strongly shaped by H~{\small I} column density \citep{Adams1972,Neufeld1990,Verhamme2006}, as well as dust content \citep{Laursen2009} and gas kinematics \citep{Bonilha1979,Ahn2002}. 
As described in Section~\ref{sec:QSO1_Lya}, the Ly$\alpha$ emission in Abell2744-QSO1 is $5-10\times$ broader than Ly$\alpha$ emission lines in sources of similar UV luminosities at $z>6$, implying gas conditions and/or Ly$\alpha$ scattering mechanisms that are not typical of star-forming galaxies.
In this section, we investigate the H~{\small I} gas properties which may explain the Ly$\alpha$ velocity profile in Abell2744-QSO1. We will discuss the implications for the gas environment of LRDs and their host galaxies in Section~\ref{sec:discussion}.

In what follows, we consider two possible scenarios for the origin of Ly$\alpha$ and investigate the gas properties implied in each case.
We assume flux cut-off blueward of line center is due to scattering in the IGM, as seen ubiquitously in $z\gtrsim6$ Ly$\alpha$ emission lines \citep[e.g.,][]{Saxena2024,Tang2024c} and expected from the high Gunn-Peterson optical depth \citep{Gunn1965} at these redshifts \citep[e.g.,][]{Bosman2022,Qin2025}. 
Thus, we truncate all models blueward of line center when fitting and comparing with the observed profile.

\subsection{Intrinsically Broad Ly$\alpha$} \label{sec:lya_broad}


\begin{figure}
\includegraphics[width=\linewidth]{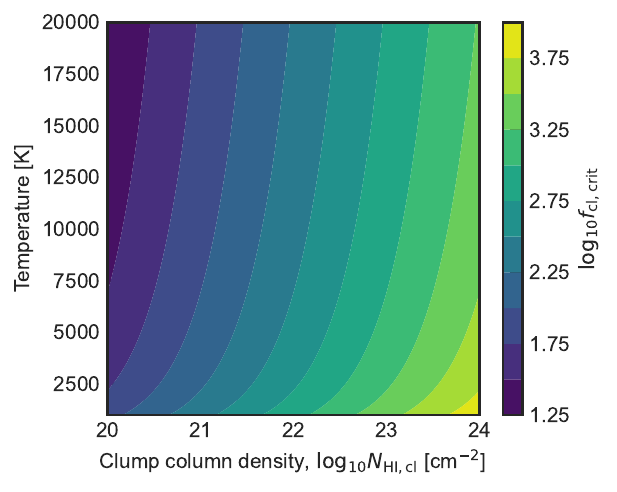}
\caption{Maximum number of dense gas clumps along the line-of-sight, $f_{\rm cl,crit}$, that would allow broad Ly$\alpha$ to escape without significant redshifting. We show $f_{\rm cl,crit}$ as a function of the clump column density and temperature, following \citet{Gronke2017a}. The large values of $f_{\rm cl,crit}$ imply that there can be a very high covering fraction ($F_{\rm cov} \lesssim 1- e^{-f_{\rm cl}}$) of dense clumps and some Ly$\alpha$ can still escape without significantly altering the lineshape.}
\label{fig:Lya_clumpy}
\end{figure}

Motivated by the similarity between the Ly$\alpha$ and H$\alpha$ profiles (Figure~\ref{fig:QSO1_lya}), we first consider a scenario where Ly$\alpha$ and the Balmer lines are produced and broadened to FWHM $\gtrsim2000$~km~s$^{-1}$ in the same region -- for example, via Doppler shifting in a BLR and/or Thomson scattering, though we note our spectrum does not reach sufficient S/N to determine if the Ly$\alpha$ shows an exponential wing. 
In this case, the intrinsic Ly$\alpha$ line profile should be similar to H$\alpha$. 
The broad lines subsequently propagate through dense neutral gas which may imprint absorption in the Balmer lines and continuum \citep[$N_{\rm HI} \sim 10^{24}$~cm$^{-2}$, e.g.,][]{DEugenio2025a,Ji2025,Naidu2025}. 

Figure~\ref{fig:Lya_tau} shows the Ly$\alpha$ spectra predicted if an intrinsically broad emission line (FWHM $=2600$~km~s$^{-1}$, assuming the broad component of H$\alpha$, see Table~\ref{tab:QSO1_opt_line}) resonantly scatters through a static, uniform shell of gas, assuming different H~{\small I} column densities and a gas temperature of $10^4$~K, using the Monte Carlo radiative transfer code \texttt{tlac} \citep{Gronke2014}.
It is clear that intervening gas with column density $N_{\rm HI} \gtrsim 10^{23}$~cm$^{-2}$, i.e., as typically required to explain the Balmer break and Balmer line absorption, would significantly alter the observed Ly$\alpha$ spectrum, as it would be extremely optically thick ($\tau > 10^3$) to photons up to $\approx1000$~km~s$^{-1}$ redward of line center.
In a static, uniform medium, Ly$\alpha$ photons are more likely to escape if they diffuse in frequency/velocity beyond these velocities; thus the profile emerging from such an optically thick medium should be significantly redshifted \citep[e.g.,][]{Adams1972,Neufeld1990}.
Outflows and/or random motions in the gas can facilitate Ly$\alpha$ escape closer to line center if photons appear redshifted away from the high optical depth in the rest-frame of the gas. 
However this would require extreme velocities (i.e., $>1000$~km~s$^{-1}$ if $N_{\rm HI} > 10^{23}$~cm$^{-2}$), which are not consistent with the $\lesssim 100$~km~s$^{-1}$ offsets of the Balmer line absorption features in Abell2744-QSO1 \citep{DEugenio2025a}.
This suggests that, if the intrinsic line profile is broad, the majority of the observed Ly$\alpha$ emission in Abell2744-QSO1 does not scatter through a uniform medium of high column density gas ($N_{\rm HI} > 10^{23}$~cm$^{-2}$).

In this scenario, the only viable way for Ly$\alpha$ to escape without significant frequency redistribution is if the dense gas is clumpy, embedded in a lower density medium, such that Ly$\alpha$ photons mostly scatter off the surface of dense clumps \citep{Neufeld1991,Hansen2006}.
This occurs if Ly$\alpha$ photons are more likely to escape by scattering via random walk \textit{between} clumps than by diffusing in frequency by resonantly scattering \textit{through} clumps.
\citet{Gronke2016} demonstrated there is a critical number of clumps, $f_{\rm cl,crit}$, along the line-of-sight, above which resonant scattering significantly alters the Ly$\alpha$ line shape. 
At the high column densities considered here, $N_{\rm HI,cl} \gtrsim 10^{20}$~cm$^{-2}$, even clumps with large random velocities will be optically thick to photons emitted at Ly$\alpha$ line center (i.e., $\tau_{{\rm Ly}\alpha}(v_{\rm cl}) > 1$, even if $v_{\rm cl}\approx1000$~km~s$^{-1}$).
In this case, the critical number of clumps is a function of the clump H~{\small I} column density, $N_{\rm HI,cl}$, and temperature:
$f_{\rm cl,crit} \approx (N_{\rm HI,cl}/10^{17}\,{\rm cm}^{-2})^{1/2} (T/10^4\,{\rm K})^{-1}$
\citep[essentially the static case described by][see their Equation 9]{Gronke2017a}.
If the number of clumps along the line of sight $f_{\rm cl} < f_{\rm cl,crit}$, it is possible for some Ly$\alpha$ photons to escape with little resonant scattering and there is no significant change in the intrinsic line shape.
The angular covering fraction of clumps can be estimated assuming an isotropic Poisson distribution of clumps: $F_{\rm cov} = 1-e^{-{f_{\rm cl}}}$.
Consequently, even a medium with a near unity covering fraction is effectively `porous' to Ly$\alpha$, providing the number of clumps per sightline is below $f_{\rm cl,crit}$.

Figure~\ref{fig:Lya_clumpy} shows $f_{\rm cl,crit}$ as a function of the clump H~{\small I} column density and gas temperature.
We show the range of H~{\small I} ($n=1$) column densities required to produce Balmer line absorption, providing the $n=2$ population is boosted by collisional excitation ($n_2/n_1 \sim 10^{-6}$), and temperatures which have been suggested for the neutral envelopes of LRDs ($\sim2000-2\times10^4$~K; e.g., \citealt{deGraaff2025c}).
Given these high column densities, we find the maximum covering fraction of clumps could be up to almost unity (e.g., $f_{\rm cl,crit} \approx 15$ for $N_{\rm HI,cl}=10^{20}$~cm$^{-2}$), and the Ly$\alpha$ line shape could still be preserved.
Increasing the column density raises $f_{\rm cl,crit}$, and correspondingly, the maximum covering fraction before which the line profile will change, as photons preferentially escape via a random walk reflecting off optically thick clumps rather than scattering through them.
Conversely, higher gas temperatures at fixed $N_{\rm HI,cl}$ decrease $f_{\rm cl,crit}$ as the clump optical depth is reduced via Doppler broadening.
While derived for Ly$\alpha$ photons close to line center, these conclusions should hold for photons at higher velocities, as the timescale for escape via reflections remains significantly shorter than escape via scattering in very optically thick clumps \citep{Gronke2017a}.
Future dedicated radiative transfer simulations will be important for understanding the impact of such dense clumps on the extended wings of Ly$\alpha$.
Overall, this implies that the Ly$\alpha$ profile can be compatible with high H~{\small I} column densities, provided the gas is clumpy, and that the clumps could have a high angular covering fraction.

Our detections of permitted Fe~{\small II} and O~{\small I} UV lines in Abell2744-QSO1 provide additional evidence for a clumpy medium, and are important for reconciling the Ly$\alpha$ profile with the low Ly$\alpha$ escape fraction ($<10\%$).
These transitions are likely excited via Ly$\alpha$/Ly$\beta$ pumping in optically thick gas \citep{Kwan1981,Sigut2003}.
In a clumpy medium, Ly$\alpha$ (and Ly$\beta$) photons will scatter in the surface of optically thick clumps before escaping \citep{Neufeld1991,Hansen2006}, providing a channel for Ly$\alpha$ destruction via fluorescence -- explaining the low $f_{\rm esc,Ly\alpha}$ -- while simultaneously providing a mechanism to pump the H~{\small I} $n=2$ population required for Balmer line absorption.
The profiles of these lines also help to localize the clumps.
The narrow widths of the Fe~{\small II} and O~{\small I} lines ($\sim300$~km~s$^{-1}$) and the similarity between Ly$\alpha$ and H$\alpha$ profiles imply the clumps are in a lower velocity, low optical depth ($\tau_e$) outer region of the LRD compared to the broad line emitting region\footnote{A high $\tau_e$ would preferentially broaden the Ly$\alpha$ wings relative to H$\alpha$ as Ly$\alpha$ photons traverse a longer path length through the medium as they scatter via a random walk off clumps \citep{Gronke2017a}.}.
Deeper grating spectroscopy of the extended wings of these lines would help to further constrain the location of the clumps.

If dust is present around the LRD and the majority of the dust is in dense clumps, Ly$\alpha$ photons could pass through the medium without significant dust absorption \citep{Neufeld1991,Laursen2013,Gronke2017a}, while the Balmer lines, and other permitted lines with lower cross-sections than Ly$\alpha$, would propagate through the clumps and be partially absorbed by dust as they experience a higher dust optical depth. 
If this is the case for Abell2744-QSO1, the observed Ly$\alpha$ escape fraction would be larger than expected from the broad line dust attenuation.
Assuming the SMC extinction law \citep{Gordon2003}, the broad line attenuation $A_{\rm H\alpha,b}=1.8\pm0.6$~mag derived from the Balmer decrement (Section~\ref{sec:QSO1_Lya}) corresponds to a Ly$\alpha$ escape fraction due to dust absorption of $f_{\rm esc,Ly\alpha} \sim e^{-\tau_{\rm dust}}<0.02$\% \citep[e.g.,][]{Verhamme2006}. 
This is over an order of magnitude lower than the Ly$\alpha$ escape fraction we estimated for the broad component ($0.7\pm0.2$\%). 
If the Balmer decrement is mostly due to dust attenuation, this implies that Ly$\alpha$ photons are less attenuated than the Balmer lines, with a clumpy medium providing a potential explanation for the higher than predicted escape fraction.
As noted in Section~\ref{sec:QSO1_Lya}, collisional excitation in dense gas may boost the Balmer decrement, and also Ly$\alpha$ flux, which may impact the interpretation of the Ly$\alpha$ escape fractions. 

In summary, dense clumps allow the transmission of broad Ly$\alpha$ without significant resonant scattering or dust attenuation. 
We discuss how such a clumpy medium could self-consistently produce the Balmer break and line absorption in Section~\ref{sec:discussion}.

\subsection{Intrinsically Narrow Ly$\alpha$} \label{sec:lya_narrow}


\begin{figure}
\includegraphics[width=\linewidth]{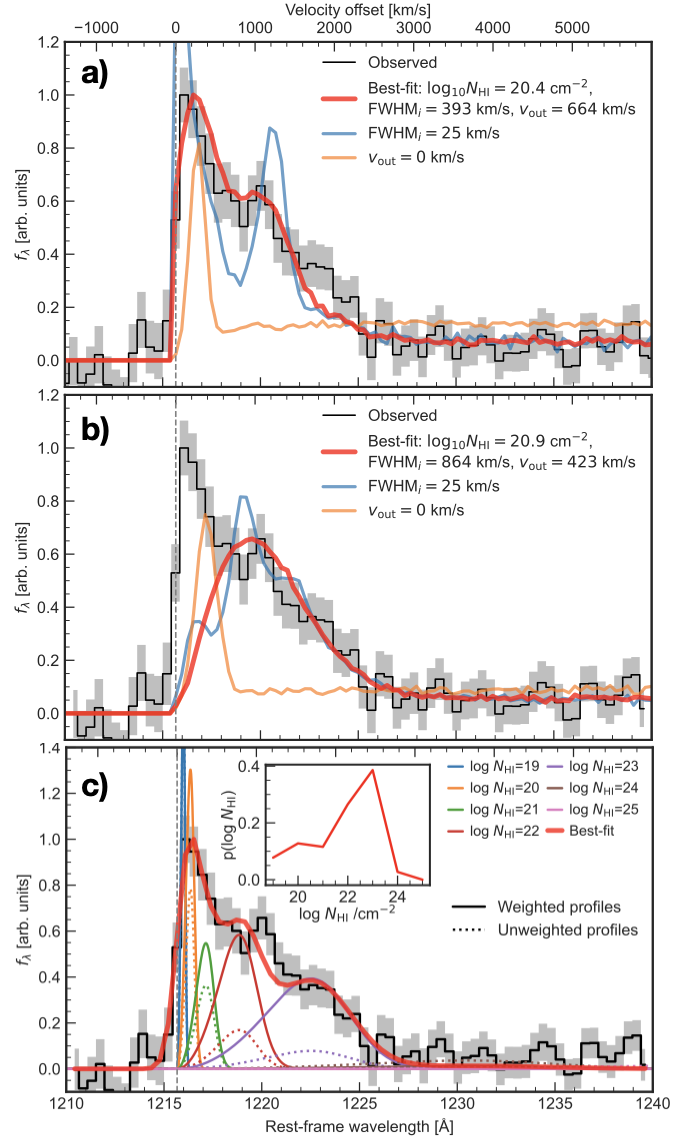}
\caption{Best-fit Ly$\alpha$ profiles for Abell2744-QSO1 (red lines), compared to the observed profile (black) assuming the profile is shaped by resonant scattering for three cases described in Section~\ref{sec:lya_narrow}. a) The \texttt{zELDA} fit to the full profile which requires high outflow velocities and a high intrinsic linewidth. Reducing the intrinsic FWHM (blue line) or outflow velocity (orange line) cannot reproduce the observed profile. b) The \texttt{zELDA} fit to the broad component of the line, which is qualitatively similar to the full profile fit but requires a higher column density. c) Assuming Ly$\alpha$ scatters through an inhomogeneous H~{\scriptsize I} distribution and thus can be described by a weighted sum of profiles at different H~{\scriptsize I} column densities \citep[see Section~\ref{sec:lya_model}, following][]{AlmadaMonter2026}. Colored lines show the weighted (solid) and unweighted (dashed) profiles at different $N_{\rm HI}$. The inset plot shows the $N_{\rm HI}$ probability distribution function recovered from the best-fit weights.}
\label{fig:Lya_fit}
\end{figure}

Alternatively, Ly$\alpha$ may originate from outside of the region where Balmer absorption occurs. 
For example, in star-forming regions in the host galaxy or in the dense nuclear region around the LRD \citep[e.g.,][]{Asada2026,Inayoshi2026}. 
In this case, the intrinsic Ly$\alpha$ emission would be narrow (broadened only by thermal and turbulent motions in H~{\small II} regions), and the observed broad profile would arise from resonant scattering in dense gas around the star-forming regions. 
The similarity between Ly$\alpha$ and H$\alpha$ in Abell2744-QSO1 would then be coincidental.

We now consider what gas conditions can produce a Ly$\alpha$ profile consistent with Abell2744-QSO1 via resonant scattering. 
We first fit the profile assuming Ly$\alpha$ scatters through a uniform gas shell with fixed $N_{\rm HI}$, including outflows \citep[e.g.,][]{Ahn2002,Verhamme2006}. 
We use the \texttt{zELDA} code \citep{GurungLopez2019,GurungLopez2022}, which is built on a large grid of Ly$\alpha$ Monte Carlo radiative transfer simulations, to fit the profile as a function of: the shell column density $N_{\rm HI}$; the outflow velocity, $v_{\rm out}$; dust optical depth $\tau_{\rm dust}$; the intrinsic linewidth (e.g., set by thermal motions in the emitting region); and the intrinsic Ly$\alpha$ EW.
We use \texttt{zELDA} to fit the observed emission line with an MCMC, using a Gaussian likelihood function.
We fit both the full profile, and the broad component (as derived in Section~\ref{sec:QSO1_Lya}) alone -- assuming the narrow component described in Section~\ref{sec:QSO1_Lya} is produced outside of the LRD region (we subtract the narrow component, which, with FWHM $\approx 300$~km~s$^{-1}$ is comparable to Ly$\alpha$ seen in star-forming galaxies; Figure~\ref{fig:lya_fwhm}). 
As above, we apply IGM attenuation to the blue side of the model. 
Our goal is to understand the range of gas properties that would be required to recreate the observed line profile, and we focus here on the inferred kinematics and column density.
We show the fitting results in Figure~\ref{fig:Lya_fit}. 

In both cases, we obtain reasonable fits, but the inferred intrinsic line widths, before scattering, are very broad (FWHM $>300$~km~s$^{-1}$), larger than the optical forbidden lines in Abell2744-QSO1 (see Table~\ref{tab:QSO1_opt_line}).
Fitting the full profile (Figure~\ref{fig:Lya_fit}a) requires an intrinsic Ly$\alpha$ linewidth before scattering FWHM$_i\approx 300-400$~km~s$^{-1}$ (68\% range), and for the Ly$\alpha$ to scatter through gas with moderate column density $N_{\rm HI} \approx 10^{20.5}$~cm$^{-2}$ outflowing\footnote{In this case the secondary red peak stems from ``backscattered'' Ly$\alpha$ photons, which thus obtain a $\sim 2 v_{\rm out}$ frequency boost \citep{Verhamme2006}.} at $v_{\rm out}\approx600-700$~km~s$^{-1}$.
Slower outflow velocities would shift the emergent Ly$\alpha$ peak to smaller velocity offsets, while narrower intrinsic lines would produce a profile more sharply peaked than the observed one, because most photons would scatter near line center rather than in the wings.
Higher column densities would broaden and shift the peak redwards beyond what is observed.
Fitting only the broad Ly$\alpha$ component yields similar conclusions (Figure~\ref{fig:Lya_fit}b). 
These fits require an extremely high intrinsic linewidth FWHM$_i\approx 400-1200$~km~s$^{-1}$, and scattering through a shell with high outflow velocity $v_{\rm out} \approx 400-500$~km~s$^{-1}$ and moderate $N_{\rm HI} \approx 10^{20.9}$~cm$^{-2}$.
Such high intrinsic line widths and outflow velocities considerably exceed those inferred from Ly$\alpha$ in $z\sim3-6$ galaxies with similar UV magnitudes to Abell2744-QSO1 \citep{Gronke2017b,Karman2017}.
The requirement for high intrinsic line widths implies highly supersonic turbulence (Mach number $M\gtrsim50$) that would be unsustainable in homogeneous neutral gas \citep[e.g.,][]{MacLow1999}, suggesting that the Ly$\alpha$ emission in Abell2744-QSO1 is not predominantly broadened by resonant scattering in a homogeneous medium.

Alternatively, the Ly$\alpha$ line profile may be explained by resonant scattering through an inhomogeneous gas distribution.
Turbulence, which has been suggested to be important for producing smooth Balmer breaks in LRDs \citep{Ji2025,Naidu2025}, has been shown to drive broad column density distributions in gas clouds \citep[e.g.,][]{Vazquez-Semadeni1998,Federrath2010}.
We fit the observed Ly$\alpha$ profile following \citet{AlmadaMonter2026} who showed Ly$\alpha$ emission from inhomogeneous H~{\small I} distributions traces the full distribution of H~{\small I} and can be approximated as a sum of analytic models at a given $N_{\rm HI}$ weighted by $p(\log_{10} N_{\rm HI})$.
Thus, we perform a maximum-likelihood fit to the observed profile using a linear combination of analytic models for Ly$\alpha$ emerging from a sphere \citep{Dijkstra2006} to estimate $p(\log_{10} N_{\rm HI})$.
We show our best-fit model in Figure~\ref{fig:Lya_fit}c, along with the recovered $p(\log_{10} N_{\rm HI})$ distribution.
To better understand what drives the recovered $p(\log_{10} N_{\rm HI})$, we also show the individual shell model profiles for a range of $N_{\rm HI}$.
We find the majority of the gas needs to be $N_{\rm HI} \sim 10^{22-23}$~cm$^{-2}$ to reproduce the emission at $\sim1000-2000$~km~s$^{-1}$, with a smaller fraction at lower and higher columns: $\sim30\%$ with $N_{\rm HI} \lesssim 10^{21}$~cm$^{-2}$, and $<10\%$ with $N_{\rm HI} \gtrsim 10^{23}$~cm$^{-2}$, consistent with our discussion of Figure~\ref{fig:Lya_tau}.
$N_{\rm HI}\sim 10^{22-23}$~cm$^{-2}$ is more than an order of magnitude higher than typical H~{\small I} column densities inferred for star-forming galaxies with similar UV luminosities to Abell2744-QSO1 at $z\gtrsim 3$ \citep[e.g.,][]{Reddy2016,Heintz2025,Mason2026,Umeda2026a}.
While this implies the broad Ly$\alpha$ is unlikely to come from the surrounding host galaxy, such high column densities could be consistent with a scenario where the Ly$\alpha$ emission is broadened by resonant scattering if it is produced in star-forming regions within a dense, inhomogeneous, nuclear disk \citep[see e.g.,][]{Asada2026,Inayoshi2026}.

To summarize, the Ly$\alpha$ profile in Abell2744-QSO1 is unusually broad given its UV magnitude (Figure~\ref{fig:lya_fwhm}). 
Given the similarity between the Ly$\alpha$ and H$\alpha$ profiles, one possibility is that Ly$\alpha$ is produced and broadened in the same region as the Balmer lines.
To maintain this similarity, we have shown the observed Ly$\alpha$ photons must primarily escape through $N_{{\rm HI},n=1} < 10^{23}$~cm$^{-2}$ gas, implying the very dense gas responsible for absorption in the Balmer lines cannot fully cover the broad line emitting region.
A multiphase medium with dense clumps may allow Ly$\alpha$ to escape without significant frequency distribution or dust attenuation, while still allowing a high covering fraction of dense gas.
Alternatively, Ly$\alpha$ could be produced and resonantly scattered outside of the region where the Balmer emission lines and Balmer absorption are produced, i.e., from a star-forming region.
In this case, we find the most viable solution to match the line profile requires intrinsically narrow Ly$\alpha$ scattering through an inhomogeneous H~{\small I} distribution peaking at $N_{\rm HI}\approx10^{22-23}$~cm$^{-2}$. 
While this is over an order of magnitude higher than typical for star-forming galaxies, it could be consistent with dense gas in the nuclear region.
Given the similarity of the Ly$\alpha$ and H$\alpha$ profiles, and that broad Ly$\alpha$ must also be produced along with the broad Balmer lines in the LRD, we consider the former case to be the most likely origin of the Ly$\alpha$ in Abell2744-QSO1.
In the next section we will discuss the implications of these results for the geometry of dense gas around LRDs.

\section{Discussion} \label{sec:discussion}

Ly$\alpha$ emission has been detected in many LRDs \citep{Ning2024,Asada2026,Torralba2026a}, including several at $z>7$ \citep{Furtak2024,Jones2026,Morishita2026}. 
However, most interpretation to date of these lines has been limited by the low resolution of NIRSpec prism spectroscopy.
Our ultra-deep G140M spectroscopy of Abell2744-QSO1, a $z=7.04$ LRD, has revealed high EW Ly$\alpha$ emission with a remarkably similar profile to the broad H$\alpha$ line.
This detection is striking on two levels.
Firstly, the Ly$\alpha$ profile is incompatible with the high H~{\small I} column density ($N_{\rm HI} \sim 10^{24}$ cm$^{-2}$) typically invoked to explain LRDs' Balmer breaks and Balmer absorption features \citep[e.g.][]{deGraaff2025b,Inayoshi2025,Ji2025,Naidu2025,Taylor2025,Sneppen2026}: Ly$\alpha$ would be redshifted to $>2000$~km~s$^{-1}$ to escape such high $N_{\rm HI}$ gas.
Secondly, due to the increasingly neutral IGM, at $z>7$ high EW Ly$\alpha$ ($\gtrsim100$~\AA) is typically only expected from sources in large ionized bubbles \citep[e.g.,][]{Lu2024,Tang2024c,Napolitano2024,Chen2025}.
Here we discuss the implications of our results for the environments of LRDs during reionization (Section~\ref{sec:igm}) and for the interpretation of dense gas around LRDs (Section~\ref{sec:geometry}).

\subsection{Ionized Regions around LRDs} \label{sec:igm}


\begin{figure}
\includegraphics[width=\linewidth]{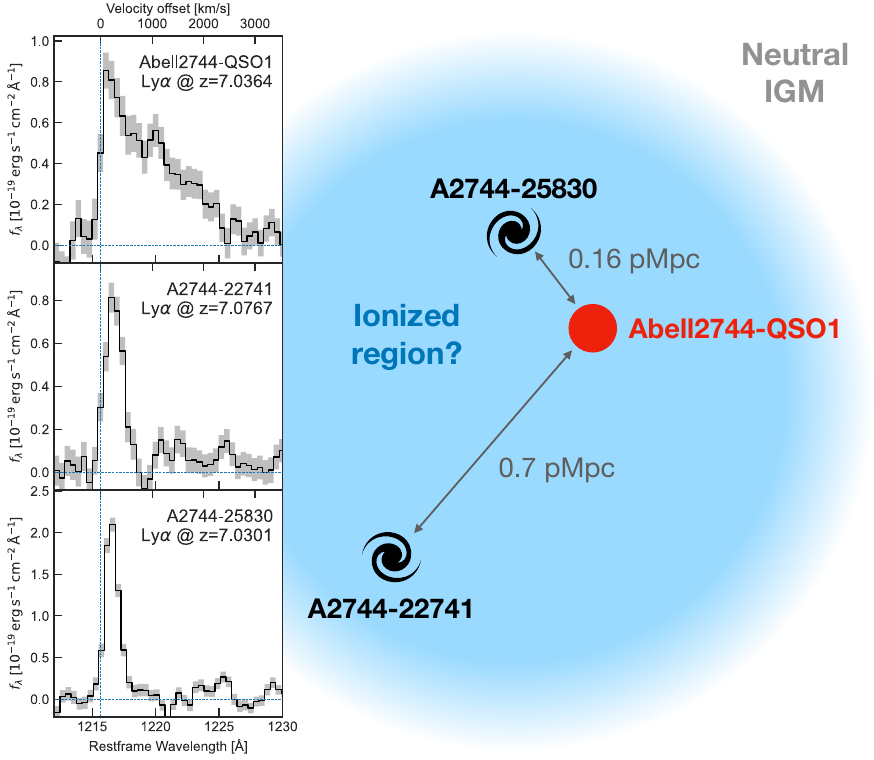}
\caption{Overview of the three Ly$\alpha$ detections at $z\approx7.04$ from Abell2744-QSO1, Abell2744-22741 and Abell2744-25830. Finding three strong Ly$\alpha$ emitters in close proximity ($\lesssim0.7$~pMpc in the source plane) at $z>7$ suggests they likely reside within an ionized region.}
\label{fig:bubble}
\end{figure}

Our deep grating spectroscopy observed Abell2744-QSO1 and its surroundings, providing insights into the environment of an LRD at a redshift where the IGM is mostly neutral \citep[e.g.,][]{Tang2024c,Kageura2025}. 
We spectroscopically confirmed two $z\approx7.04$ galaxies within $0.68$~pMpc (in the source plane) of Abell2744-QSO1, indicating the LRD may trace a dense environment at these redshifts. 
We illustrate these detections in Figure~\ref{fig:bubble}.
This is the first evidence that Abell2744-QSO1 has neighbors. 
Previous spectroscopy in Abell 2744 has been incomplete at this redshift: e.g., at $z>7$ [O~{\small III}]+H$\beta$ falls out of the range of the F356W grism used in All the Little Things (ALT, GO 3516, PIs: J. Matthee, R. Naidu; \citealt{Naidu2024}), and no other $z=7.04$ sources were confirmed in UNCOVER MSAs \citep{Bezanson2024}. 
While more complete spectroscopy in Abell 2744 will be needed to verify this, the SPURS detections indicate a significant overdensity.
In particular, Abell2744-QSO1 and A2744-25830 are separated by only $0.16$~pMpc in the source plane and $\Delta z = 0.0063$. 
Based on the \citet{Bouwens2021} $z \approx 7$ UV luminosity function, the expected number of M$_{\rm UV} < -16.9$ galaxies in a cylindrical volume of this scale is just $N \approx 0.02$. 
This is consistent with recent findings that LRDs preferentially reside in overdense environments: \citet{Matthee2025} reported LRDs reside in regions $\sim 6\times$ overdense on $\sim 0.1$~pMpc scales \citep[see also, e.g.,][]{Fujimoto2024,Labbe2024,Schindler2025,Morishita2026}.

Most notably, our spectra reveal that both neighboring galaxies also show Ly$\alpha$ emission, as well as signatures of intense radiation fields (Section~\ref{sec:z7_lae}).
All three sources have Ly$\alpha$ EWs considerably exceeding the median at $z=6.5-8.0$ \citep[$\approx5$~\AA,][]{Tang2024c}.
Finding three strong Ly$\alpha$ emitters in close proximity at $z>7$ is very rare, as the neutral IGM suppresses Ly$\alpha$ unless sources sit in ionized regions \citep[e.g.,][]{Napolitano2024,Tang2024c,Witstok2024,Chen2025,Kageura2025}.
Additionally, all three show Ly$\alpha$ emission with flux close to systemic velocity (Figure~\ref{fig:bubble}). 
As the damping wing optical depth from the neutral IGM preferentially attenuates flux close to Ly$\alpha$ line center, this further points to an ionized region around the galaxies \citep[e.g.,][]{MiraldaEscude1998,Dijkstra2007}.
In particular, the Ly$\alpha$ emission in Abell2744-25830 is offset by only $+123\pm92$~km~s$^{-1}$, one of the lowest known offsets at $z>7$ \citep{Saxena2023,Witstok2024,Tang2024b,Tang2024a}. 
Detailed analysis would required higher resolution spectroscopy, however, this may indicate the source resides in an ionized region with low residual neutral fraction which may be expected in the presence of hard radiation fields \citep{Mason2020}.
We obtain an initial estimate of the size of the ionized region, we use the Ly$\alpha$ EW, and 1$\sigma$ uncertainties, of the three sources to calculate the median fraction of Ly$\alpha$ transmitted through the IGM, relative to $z\sim5-6$ galaxies, following the approach of \citet{Tang2024c}, finding $T_{\rm IGM}>0.56$ (95\% lower limit). 
We compare this to the fraction of Ly$\alpha$ flux transmitted through along ionized sightlines with a range of sizes, calculated using the IGM damping wing at $z=7$, following the approaches in \citep{Mason2020,Endsley2022b,Prieto-Lyon2023}, assuming Gaussian emission lines centered at the velocity offset of the two new Ly$\alpha$ detections.
We find this implies an ionized region $\gtrsim 1$~pMpc \citep{Endsley2022b,Prieto-Lyon2023}, which is comparable to the median sizes of ionized bubbles predicted by simulations at these redshifts \citep[e.g.,][]{Lu2024,Neyer2024}. 
More complete spectroscopy to confirm other $z\approx7$ sources in this region and measure their Ly$\alpha$ emission will enable improved constraints on the bubble size \citep[e.g.,][]{Nikolic2025}.

These Ly$\alpha$ detections raise questions regarding the role of LRDs in ionizing their surroundings. 
The three sources presented here are all UV-faint ($-19 \lesssim {\rm M}_{\rm UV} \lesssim -17$), and even with optimistic assumptions on ionizing production and escape, star formation in these sources alone would not be sufficient produce such a $R>1$~pMpc ionized region \citep[e.g.,][]{Mason2020}.
On one hand, it is possible that LRDs simply trace overdensities, as we have noted above. 
But the Ly$\alpha$ visibility around such faint systems could also be explained if the a system like Abell2744-QSO1 contributed 
significantly to ionizing its surroundings, potentially in a past (pre-LRD) phase.

To assess whether LRDs preferentially trace ionized regions we estimate the redshift evolution in Ly$\alpha$ visibility (quantified as the fraction of sources with Ly$\alpha$ EW $>25$~\AA; e.g., \citealt{Stark2011}) of LRDs.
We measure the Ly$\alpha$ EWs (or $3\sigma$ upper limits if non-detection) of 87 LRDs at $z>4$ with spectra covering Ly$\alpha$, from the \citet{deGraaff2025c} LRD catalog, using the publicly-available prism spectra reduced by DJA \citep{deGraaff2025a,Heintz2025}.
Following the approach in \citet{Tang2024c}, we derive the Ly$\alpha$ EW distributions of LRDs in two redshift bins: $z=4-6$ and $z>6$. 
We find that the Ly$\alpha$ fractions of LRDs are consistent with no significant evolution between $z=4-6$ ($31^{+6}_{-6}\%$) and $z>6$ ($24^{+10}_{-8}\%$), in contrast with a factor of $\sim2\times$ decrease in the Ly$\alpha$ fraction in star-forming galaxies over the same redshift range \citep[e.g.,][]{Schenker2014,Pentericci2018,Mason2018,Nakane2024,Napolitano2024,Tang2024c}. 

While more precise population estimates will require larger samples with sensitive grating spectra to measure low EW emission, this result suggests that the Ly$\alpha$ emission of LRDs may be less attenuated by the neutral IGM than typical star forming galaxies, as would be expected if LRDs tend to trace larger ionized regions.
Future work to establish the prevalence of Ly$\alpha$ in LRDs and galaxies in their surroundings at $z\gtrsim7$, relative to regions without LRDs \citep[e.g.,][]{Chen2025}, will be essential for assessing their impact on the neutral IGM. 

\subsection{Implications for LRD Structure from Deep UV Spectroscopy} \label{sec:geometry}

\begin{figure}
\includegraphics[width=\linewidth]{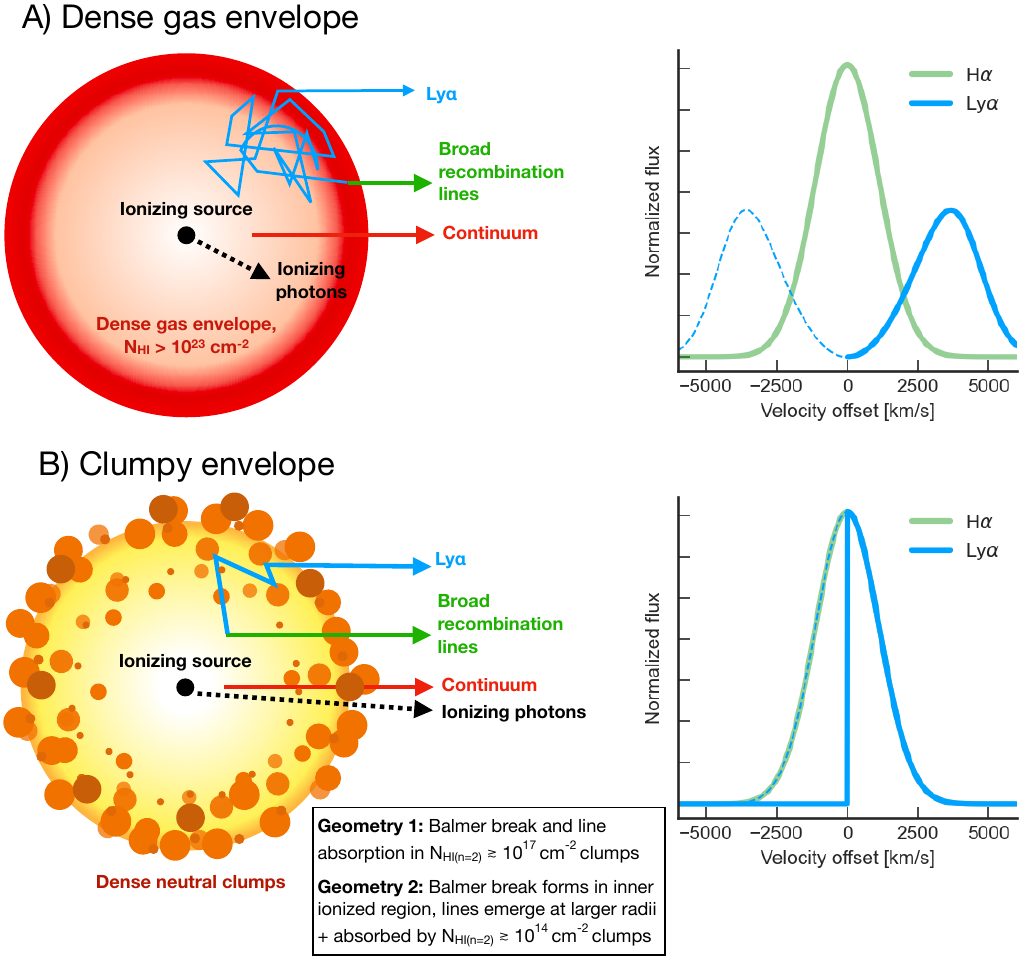}
\caption{Schematic illustrating the potential geometry of Abell2744-QSO1. A) In a uniform dense gas envelope invoked to form the Balmer break and line absorption in LRDs, Ly$\alpha$ would be extremely redshifted and broadened line relative to H$\alpha$ (see Section~\ref{sec:lya_broad}). B) A high covering fraction of dense neutral clumps around the line emitting region may imprint absorption on the Balmer lines, while some Ly$\alpha$ photons can escape by reflecting off the clumps, preserving the similar Ly$\alpha$ and H$\alpha$ line profiles.}
\label{fig:cartoon}
\end{figure}

Rest-frame UV grating spectroscopy is now starting to provide new insights into the structure of LRDs and the origin of their UV emission.
In particular, spectrally resolving Ly$\alpha$ provides a critical test for the dense gas picture posited to explain LRDs' optical spectra \citep[see also,][]{Torralba2026a}.
Dense, high $N_{\rm HI}$ gas is required to provide the H~{\small I} $n=2$ population necessary to form a strong Balmer break ($N_{{\rm HI},n=2}\gtrsim 10^{17}$~cm$^{-2}$).
However, the Ly$\alpha$ profile we have observed in Abell2744-QSO1 demonstrates that not all photons experience such high H~{\small I} columns. 
Reconciling this requires adjustments to the dense gas picture.
Motivated by our Ly$\alpha$ results, we discuss two alternative geometries to explain the spectral features of Abell2744-QSO1, and discuss how this may generalize to other LRDs. 
We illustrate these geometries, relative to a uniform dense gas picture, in Figure~\ref{fig:cartoon}.


One natural solution to explain the Ly$\alpha$ is that the dense neutral envelopes invoked for LRDs are clumpy, as we demonstrated in Section~\ref{sec:lya_broad}.
In this picture (Geometry B1 in Figure~\ref{fig:cartoon}), the incident continuum and broad emission lines propagate through a layer of dense neutral clumps. 
Ly$\alpha$ photons escape without significant frequency distribution by scattering off the surfaces of the clumps \citep[e.g.,][]{Neufeld1991,Hansen2006,Gronke2016,Chang2023}, while photons with lower H~{\small I} absorption cross-sections transmit through the clumps, where they can be absorbed by H~{\small I} in the $n=2$ state, forming the Balmer break and line absorption.
This is similar to a broad line region with a large covering fraction of clouds \citep{Inayoshi2025}.
This interpretation may also help address one of the challenges of the dense gas envelope picture: self-consistently explaining LRDs' smooth reddened Balmer breaks \citep{deGraaff2025a,Ji2025,Naidu2025}.
A radial velocity gradient of clumps in virial motion may help produce a smooth break, by scattering photons at a range of velocities, as has been suggested to explain the exponential-like wings of the emission lines \citep[][]{Scholtz2026}. 
A quantitative analysis of this possibility would require more detailed radiative transfer modeling, accounting for multiphase gas kinematics and Balmer line infilling, and is left for future theoretical work.

Another possibility is that the Balmer break forms {\it interior} to where the emission lines emerge (Geometry B2 in Figure~\ref{fig:cartoon}): within the dense, ionized region proposed to explain LRDs' exponential line profiles \citep{Chang2026,Rusakov2026,Torralba2026b}.
At very high electron densities ($n_e \gtrsim 10^{11}$~cm$^{-3}$), collisional excitation, as well as Ly$\alpha$ pumping, can boost the H~{\small I} $n=2$ population \citep{Dijkstra2016}, enabling a Balmer break to form even within highly ionized gas (\citealt{Begelman2026}; Chang et al. in prep; Katz et al. in prep). 
Motivated by luminous blue variable stars (LBVs) winds \citep[e.g.,][]{Humphreys1994}, which have been noted to share many spectral features with LRDs \citep[e.g.,][]{Matthee2026}, 
a radial density distribution in this ionized region could result in processes occurring in different layers.
If the break forms in a dense inner region, extreme electron scattering ($\tau_e \gtrsim 10$) would smooth the entire spectrum, which could naturally explain the smooth, reddened Balmer breaks seen in some LRDs (Katz et al. in prep), and the lack of variability \citep{Sneppen2026}, while the observed broad lines emerge from outer, lower-opacity layers ($\tau_e \sim 1-3$, consistent with line profiles in LRDs; \citealt{Matthee2026,Rusakov2026}).
To form Balmer line absorption, the line-emitting region is likely surrounded by dense clumps, similar to the previous scenario.
Notably, however, this geometry drastically relaxes the $N_{\rm HI}$ requirements for the line absorbing gas. 
As the Balmer line absorption cross-section is $\sim10^3\times$ higher than the bound-free cross-section, the line absorbing gas only requires $N_{\rm HI} \gtrsim 10^{20}$~cm$^{-2}$, which may also facilitate Ly$\alpha$ escape closer to line center. 
We note column density is a lower limit assuming the $n=2$ population in this region can also be boosted by collisional excitation and Ly$\alpha$ pumping. 
This picture could be tested by systematically comparing the $n=2$ columns inferred from lines versus those from the break in larger LRD samples.
A significant implication of this stratified ionized geometry is that if broad line profiles predominantly trace gas kinematics and $\tau_e$ at the edge of the line-emitting region, standard virial relations \citep[e.g.,][]{Greene2005,Reines2015} may introduce additional uncertainties in black hole mass estimates.

While we detected broad Ly$\alpha$ in Abell2744-QSO1, we detected no other broad permitted UV lines (N~{\small V}, C~{\small IV}, He~{\small II}, Mg~{\small II}). 
It remains an outstanding question why these lines have been so-far undetected in LRDs \citep{Lambrides2024,Tang2025}, despite being ubiquitous in broad-line AGN at lower redshifts.
One possibility, in the context of the stratified ionized geometry discussed above, is that high ionization lines 
form at radii with high $\tau_e$, and are broadened beyond detectability. 
It is unclear why the other broad lines would be suppressed in the clumpy envelope picture.
A softer ionizing SED than typical AGN could contribute to the lack of broad high ionization lines \citep[e.g.,][]{Madau2024,Wang2025b}. 
However, the non-detection of broad Mg~{\small II} (ionization potential, I.P. $=7.6$~eV) remains particularly puzzling in both scenarios, and we discuss some possible explanations.
One possibility is that, as Mg~{\small II} primarily collisionally excited, it may be intrinsically weak if the required ionization and density conditions to form the line \citep[e.g.,][]{Korista1997} exist only in a thin layer.
Furthermore, a high $n=2$ H~{\small I} population in the gas around the LRD could preferentially absorb Mg~{\small II} (4.4~eV) relative to Ly$\alpha$ (10.2~eV) via photoionization (I.P. $=3.4$~eV for $n=2$ H~{\small I}).
If the gas around Abell2744-QSO1 is metal poor\footnote{While our Fe~{\scriptsize II} detection indicates metals are present, a high incident Ly$\alpha$ flux may boost the Fe~{\scriptsize II} line strength even at low abundances.} this would also weaken the Mg~{\small II} luminosity \citep[see also,][]{Maiolino2025b}.
Finally, any dust present in the clumps would preferentially attenuate UV metal lines relative to Ly$\alpha$ as the UV lines see a higher dust optical depth through the clumps \citep{Neufeld1991,Chang2024}.
While Mg~{\small II} is a resonant line, its optical depth in the clumps is significantly lower than Ly$\alpha$ (e.g., $\sim10^{8}\times$ lower given the estimated metallicity of Abell2744-QSO1), making it more likely to scatter through the clumps compared to Ly$\alpha$ \citep{Gronke2017a}. 
Comparisons of broad Mg~{\small II}, if detected, and Ly$\alpha$ lines may thus be a sensitive probe of the clump covering fraction \citep{Chang2024}.
Deeper spectroscopy of Mg~{\small II} in LRDs promises to place stronger limits on broad emission and test these scenarios.

Our detection of a narrow high ionization line, [Ne~{\small IV}] (I.P. $>64$~eV) in UNCOVER-2476 (Section~\ref{sec:2476_high_ion}) adds to the sample of such lines detected in LRDs \citep{Maiolino2024a,Tang2025}, potentially implying the presence of radiation fields harder than stellar populations.
Whether these high ionization lines indicate AGN photoionization or other sources of hard photons around LRDs remains unclear.
Given our results,
one possibility is that random motions and outflows of the clumps, potentially aided by radiation pressure from Ly$\alpha$ trapping in their surfaces, may drive clump collisions.
Collisions could drive fast radiative shocks capable of powering high ionization lines without a hard incident spectrum \citep[e.g.,][]{Allen2008,Izotov2012,Alarie2019}.
A clumpy medium could also allow the escape of hard continuum photons along low-opacity sightlines in LRDs with non-unity covering fractions \citep[as suggested by][]{Lambrides2025,Tang2025}.
While current spectroscopy for most LRDs lacks the depth to detect high ionization UV lines \citep[typical EW $\lesssim 10$~\AA; e.g.,][]{Chisholm2024,Tang2025}, on-going deep grating programs such as SPURS and Deep Insights into UV Spectroscopy at the Epoch of Reionization (DIVER, GO-8018, PI: X. Lin) will enable us to better assess the prevalence and origin of high ionization lines in LRDs.

Of course, the LRDs discussed in this work are only a subset of the population, and whether their diverse spectra can be explained within a unified framework remains to be seen \citep[e.g.,][]{deGraaff2025c,Madau2026,Matthee2026,Sun2026}.
Ly$\alpha$ provides a promising probe. 
$39\pm12\%$ of LRDs show Ly$\alpha$ emission in prism spectra \citep{Asada2026}, however only a handful currently have sufficient resolution to measure their line profiles.
One question is whether we expect to see broad Ly$\alpha$ in all LRDs if their outer layers are clumpy.
\citet{Morishita2026} reported broad Ly$\alpha$ in CANUCS-LRD-z8.6, with a line profile similar to Abell2744-QSO1 and also indicative of a large ionized region \citep[though that source does not have a strong Balmer break,][]{Tripodi2025}.
However, broad Ly$\alpha$ has not yet been reported in other LRDs with high resolution spectra \citep{Tang2025,Torralba2026a}.
A range of clump covering fractions around LRDs may naturally explain this diversity: in LRDs with higher clump covering fractions than Abell2744-QSO1, broad Ly$\alpha$ line profiles may be scattered beyond our current detection limits \citep[see][]{Torralba2026a}. 
Higher covering fractions may also favor the production of fluorescent lines and increase Balmer line absorption, a trend that could be tested systematically in larger samples.
Furthermore, the detection of broad Ly$\alpha$ in Abell2744-QSO1 may be facilitated by its weak host galaxy contribution: having among the lowest narrow emission line EW (e.g., [O~{\small III}]~$\lambda5007$ EW $=5$~\AA), similar to MoM-BH* \citep[3~\AA;][]{Naidu2025} and the Cliff \citep[6~\AA;][]{deGraaff2025b}.
In LRDs with a stronger host component, dust within the host galaxy may attenuate some of the broad Ly$\alpha$ flux, a hypothesis that requires a systematic comparison of Ly$\alpha$ emission and inferred dust attenuation.

While rest-frame optical spectra have revealed the extreme densities of LRDs, the rest-frame UV is providing new insights into their potential role in reionization and probes of their structure.
Our results suggest LRDs may preferentially reside in large ionized regions, and that the dense gas around them is likely clumpy.
Progress will require larger samples with deep UV grating spectroscopy to establish the visibility of Ly$\alpha$ in LRDs and their surroundings during the Epoch of Reionization, and to
measure empirical trends between Ly$\alpha$ profiles and LRDs' other spectral features. 
These observations must be supported by improved radiative transfer simulations of clumpy, high-density environments to help link empirical trends to geometries.
Future observations should also be able to detect photospheric lines in UV bright LRDs, providing constraints on the stellar contribution to the UV emission.

\section{Summary} \label{sec:summary}

We present ultra-deep (29 hours) JWST/NIRSpec G140M (rest-frame UV) spectroscopy of two LRDs in the Abell 2744 field: Abell2744-QSO1 at $z=7.0364$ and a newly-confirmed LRD UNCOVER-2476 at $z=4.0197$. 
The data were obtained as part of the SPURS Cycle 4 Large Program. 
The deep $R\simeq1000$ rest-frame UV spectra provide robust constraints on the Ly$\alpha$ velocity profile of Abell2744-QSO1 and high ionization UV emission lines (N~{\small V}, C~{\small IV}, He~{\small II}, [Ne~{\small IV}], [Ne~{\small V}]) of these two LRDs. 
With this dataset, we investigate the gas properties of LRDs and their host galaxies, as well as the galaxy environment that LRDs reside. 
We summarize our key results below.

1. With deep medium-resolution spectrum, we characterize the Ly$\alpha$ velocity profile of Abell2744-QSO1. 
The Ly$\alpha$ profile appears to be the superposition of a narrow host-like component (FWHM $=333$~km~s$^{-1}$) offset by $+258$~km~s$^{-1}$ from line center, and a broad component (FWHM $=1498$~km~s$^{-1}$) offset by $+1015$~km~s$^{-1}$. 
The FWHM of broad Ly$\alpha$ is $5-10\times$ larger than that of galaxies with similar UV luminosities (M$_{\rm UV}=-16.9$) at $z>6$, suggesting that the Ly$\alpha$ powering and scattering mechanisms are not typical of star-forming galaxies. 
The broad Ly$\alpha$ line width is similar to that of the red-side wing of the broad H$\alpha$ line, indicating that the broad Ly$\alpha$ may be linked to the mechanism that is producing the broad Balmer lines. 

2. We detect narrow C~{\small IV} emission (EW $=5.7\pm1.3$~\AA) in the SPURS G140M spectrum of Abell2744-QSO1. 
We do not find broad C~{\small IV}, He~{\small II}, or Mg~{\small II} emission. 
The $3\sigma$ upper limits on the ratios between these broad line fluxes and broad H$\beta$ flux are well below that of typical type I AGN. 
No other very high ionization UV emission line (N~{\small V}, [Ne~{\small IV}], [Ne~{\small V}]) is seen in the spectrum of Abell2744-QSO1. 
We additionally detect narrow O~{\small I}~$\lambda1302$ (EW $=3.6\pm1.1$~\AA) and Fe~{\small II}~$\lambda1786$ emission (EW $=5.7\pm1.4$~\AA). 
As O~{\small I} and Fe~{\small II} can be enhanced by Ly$\beta$ and Ly$\alpha$ fluorescence, these detections indicate that some sightlines are highly optically thick to Ly$\alpha$ and Ly$\beta$. 

3. The SPURS G140M spectrum of UNCOVER-2476 reveals narrow high ionization emission lines [Ne~{\small IV}]~$\lambda\lambda2422,2424$ (EW $=1.3\pm0.3$~\AA), He~{\small II}~$\lambda2733$, and [Fe~{\small IV}]~$\lambda\lambda2829,2835$. 
This may indicate the presence of hard photons ($>64$~eV) which are able to escape along low-opacity sightlines or fast-radiative shocks. 

4. We investigate the H~{\small I} gas properties that can explain the Ly$\alpha$ velocity profile of Abell2744-QSO1. 
We first consider if Ly$\alpha$ is produced and broadened in the same regions as the broad Balmer lines. 
If Ly$\alpha$ photons transfer through the very dense neutral gas predicted to be responsible for the Balmer Break ($N_{\rm HI}\sim10^{24}$~cm$^{-2}$), the net line profile will be significantly more redshifted and broadened than we observe. 
We argue that the Ly$\alpha$ profile can be best explained if the dense neutral gas is clumpy.
This allows Ly$\alpha$ to escape by scattering off of the clump surfaces without significant resonant scattering or dust attenuation. 

5. We newly-identify two close neighbors of Abell2744-QSO1.
One of the galaxies (Abell2744-25830) is $0.16$~pMpc away and a further source (Abell2744-22741) is $0.68$~pMpc away from Abell2744-QSO1 in source plane, both of which also show Ly$\alpha$ emission. 
These systems appear to suggest that Abell2744-QSO1 traces a dense environment, as is commonly seen in LRDs. 
The overdensity may carve out an early ionized region, contributing to the boosted Ly$\alpha$ visibility. 
Future observations are required to test if the the Ly$\alpha$ visibility of LRDs tends to be enhanced owing to the larger-than-average galaxy densities they trace.

\begin{acknowledgments}

MT acknowledges support by a Shanghai Jiao Tong University start-up grant. 
DPS acknowledges support by the National Science Foundation under Grant No. AST-2109066.
CAM acknowledges support by the European Union ERC grant RISES (101163035), Carlsberg Foundation (CF22-1322), and VILLUM FONDEN (37459).
ZC acknowledges support by VILLUM FONDEN (37459). 
The Cosmic Dawn Center (DAWN) is funded by the Danish National Research Foundation under grant DNRF140.
MG thanks the Max Planck Society for support through the Max Planck Research Group, and the European Union for support through ERC-2024-STG 101165038 (ReMMU).
LJF and RE acknowledge support from the University of Texas at Austin Cosmic Frontier Center.
SJC acknowledges support from the ERC synergy grant 101166930 – RECAP.
JM acknowledges funding by the European Union (ERC, AGENTS, 101076224).
LW acknowledges support from the Gavin Boyle Fellowship at the Kavli Institute for Cosmology, Cambridge and from the Kavli Foundation. 
AZ acknowledges support by the Israel Science Foundation Grant No. 864/23.
VG acknowledges support by the Carlsberg Foundation (CF22-1322).

This work is based on observations made with the NASA/ESA/CSA James Webb Space Telescope. 
The data were obtained from the Mikulski Archive for Space Telescopes at the Space Telescope Science Institute, which is operated by the Association of Universities for Research in Astronomy, Inc., under NASA contract NAS 5-03127 for JWST. 
These observations are associated with the program GO 9214. 
We thank our program coordinator, Christian Soto, and our NIRSpec reviewer, Diane Karakla.
The authors acknowledge the UNCOVER team led by Ivo Labb\'{e} and Rachel Bezanson for developing their observing programs with zero-exclusive-access periods.
Part of the data products presented herein were retrieved from the Dawn JWST Archive (DJA). 
DJA is an initiative of the Cosmic Dawn Center, which is funded by the Danish National Research Foundation under grant DNRF140.
The Tycho supercomputer hosted at the SCIENCE HPC center at the University of Copenhagen was used for supporting this work.

\end{acknowledgments}


%

\software{\texttt{NumPy} \citep{Harris2020}, \texttt{Matplotlib} \citep{Hunter2007}, \texttt{SciPy} \citep{Virtanen2020}, \texttt{Astropy} \citep{AstropyCollaboration2022}, \texttt{PyNeb} \citep{Luridiana2015}, \texttt{msaexp} \citep{Brammer2023}, \texttt{zELDA} code \citep{GurungLopez2019,GurungLopez2022}, \texttt{tlac} \citep{Gronke2014}}


\appendix

\restartappendixnumbering

\section{Rest-Frame Optical and NIR Spectra of UNCOVER-2476} \label{sec:2476_opt_nir_spec}


\begin{figure*}
\includegraphics[width=\linewidth]{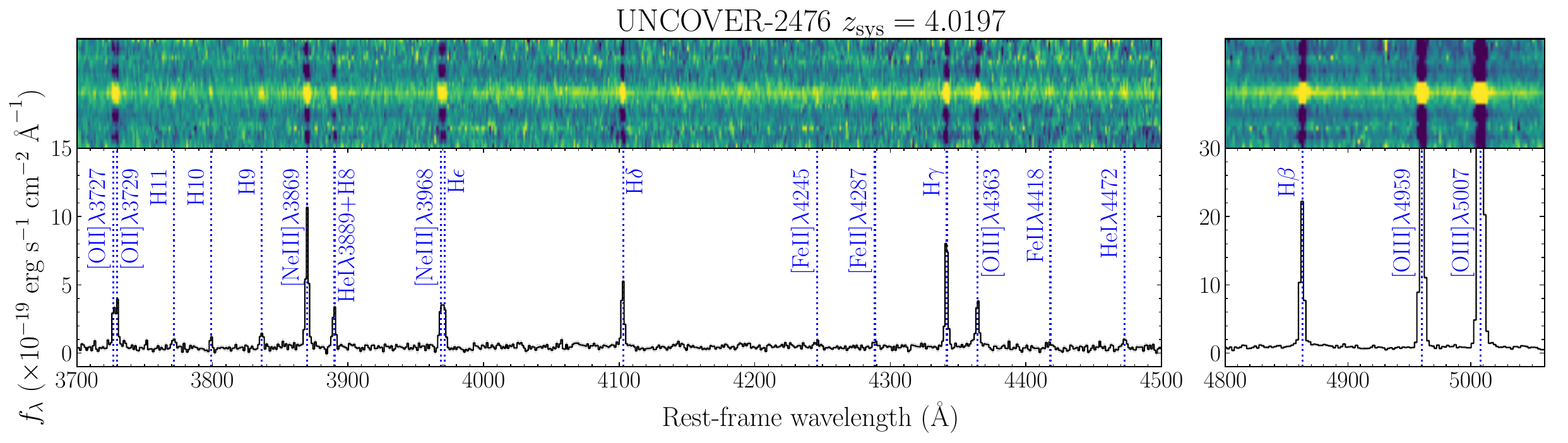}
\includegraphics[width=\linewidth]{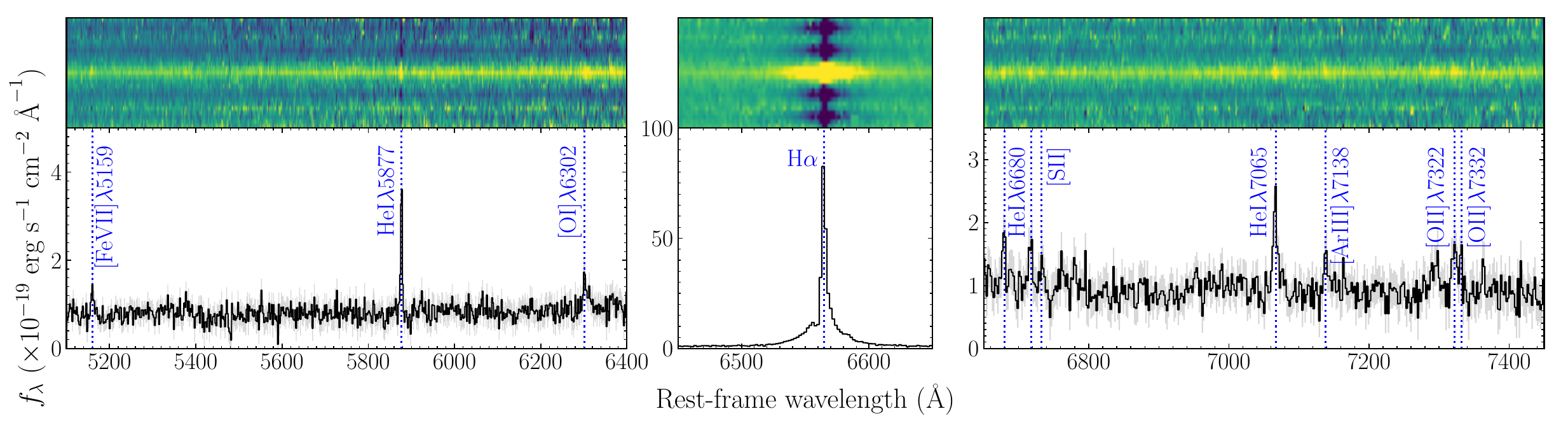}
\includegraphics[width=\linewidth]{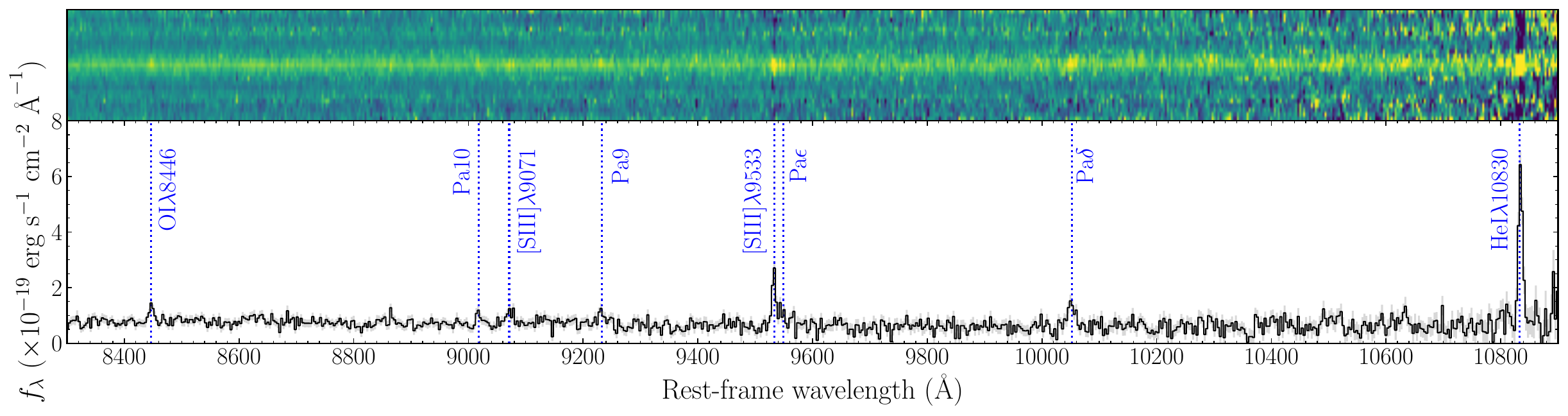}
\caption{SPURS JWST/NIRSpec rest-frame optical to NIR (G235M/F170LP and G395M/F290LP) spectra of UNCOVER-2476, shown in the same way as Figure~\ref{fig:QSO1_uv_spec}. We detect broad H$\beta$ and H$\alpha$ emission lines and several hydrogen and helium lines, as well as a suite of narrow, strong forbidden lines.}
\label{fig:2476_opt_nir_spec}
\end{figure*}

In this section, we discuss the SPURS rest-frame optical and NIR (G235M and G395M) spectra of UNCOVER-2476 (Figure~\ref{fig:2476_opt_nir_spec}) in more detail. 
We identify a suite of strong emission lines ([O~{\small II}], [Ne~{\small III}], H$\gamma$, H$\beta$, [O~{\small III}], H$\alpha$; Table~\ref{tab:2476_opt_nir_line}). 
We additionally detect several hydrogen (H11, H10, H9, H8, H$\epsilon$, H$\delta$, Pa10, Pa9, Pa$\epsilon$, Pa$\delta$) and helium (He~{\small I}~$\lambda3889$, $\lambda4472$, $\lambda5877$, $\lambda6680$, $\lambda7065$, $\lambda7281$, $\lambda10830$) emission lines. 
We also find sulfur ([S~{\small II}]~$\lambda6717$, S/N $=3$; [S~{\small II}]~$\lambda6732$, S/N $=1.5$; [S~{\small III}]~$\lambda9071$, S/N $=5$; [S~{\small III}]~$\lambda9533$ S/N $=6$) and [Ar~{\small III}]~$\lambda7138$ (S/N $=3$) emission lines.
We detect [O~{\small III}]~$\lambda4363$ (S/N $=10$), [O~{\small II}]~$\lambda7322$ (S/N $=2$), and [O~{\small II}]~$\lambda7332$ (S/N $=2$) auroral lines. 
We also report neutral oxygen lines ([O~{\small I}]~$\lambda6302$, S/N $=4$; O~{\small I}~$\lambda8446$, S/N $=3$) and iron emission lines ([Fe~{\small II}]~$\lambda4287$, S/N $=3.5$; Fe~{\small II}~$\lambda4418$, S/N $=3$; [Fe~{\small VII}]~$\lambda5159$, S/N $=4$), which are often seen in type I AGN and LRDs \citep[e.g.,][]{DEugenio2025b,Kokorev2025,Tang2025,Lin2026}. 

Both the H$\beta$ and H$\alpha$ emission lines show broad and narrow components. 
We simultaneously fit the narrow and broad H$\beta$ line profile with two Gaussians. 
We derive FWHMs of $205\pm4$~km~s$^{-1}$ (narrow) and $2123\pm496$~km~s$^{-1}$ (broad). 
The H$\alpha$ emission line additionally presents an absorption feature, which is shown in a subset of LRDs \citep[e.g.,][]{Lin2024,Matthee2024,Kocevski2025}. 
We fit the H$\alpha$ line profile with three Gaussians. 
For the broad component, we derive a FWHM of $1703\pm28$~km~s$^{-1}$, consistent with the line width of the broad H$\beta$ emission. 
The H$\alpha$ absorption line is blueshifted ($-257\pm92$~km~s$^{-1}$), with an EW of $=-6.6\pm0.9$~\AA. 

The narrow line detections allow us to characterize the properties of the narrow line emitting gas in UNCOVER-2476. 
Interpreting the narrow lines requires knowledge of the dust attenuation to these lines, which can be estimated from the Balmer decrement measurement. 
We measure a narrow H$\alpha$/H$\beta$ flux ratio of $3.37\pm0.19$. 
Assuming the SMC extinction law \citep{Gordon2003} and case B recombination \citep{Osterbrock2006}, we derive a modest narrow line attenuation of $A_V=0.54\pm0.03$~mag. 
In the following, we will correct the narrow line fluxes with this reddening value. 


\begin{deluxetable}{cccc}[h]
\tablecaption{Rest-frame optical and NIR emission line flux ($\times10^{-20}$~erg~s$^{-1}$~cm$^{-2}$), EW (\AA), and FWHM (km~s$^{-1}$) of UNCOVER-2476 measured from SPURS spectra.}
\tablehead{
Line & Flux & EW & FWHM
}
\startdata
{[}O~{\scriptsize II}]~$\lambda3727$ & $71.9\pm7.7$ & $16.4\pm1.8$ & $162\pm12$ \\
{[}O~{\scriptsize II}]~$\lambda3729$ & $77.7\pm7.7$ & $17.8\pm1.8$ & $159\pm12$ \\
H11 & $23.6\pm4.2$ & $6.7\pm1.2$ & $268\pm62$ \\
H10 & $16.2\pm3.5$ & $4.3\pm0.9$ & $217\pm35$ \\
H9 & $33.6\pm4.4$ & $8.0\pm1.0$ & $217\pm35$ \\
{[}Ne~{\scriptsize III}]~$\lambda3869$ & $252.5\pm7.4$ & $51.3\pm1.5$ & $183\pm4$ \\
He~{\scriptsize I}~$\lambda3889$+H8 & $79.4\pm7.3$ & $16.6\pm1.5$ & - \\
{[}Ne~{\scriptsize III}]~$\lambda3968$+H$\epsilon$ & $162.5\pm9.9$ & $38.0\pm2.3$ & - \\
H$\delta$ & $230.7\pm15.5$ & $33.8\pm2.3$ & $295\pm14$ \\
{[}Fe~{\scriptsize II}]~$\lambda4287$ & $16.6\pm4.7$ & $3.9\pm1.1$ & $245\pm66$ \\
H$\gamma$ & $400.5\pm18.5$ & $56.9\pm2.6$ & $275\pm9$ \\
{[}O~{\scriptsize III}]~$\lambda4363$ & $183.7\pm17.5$ & $26.0\pm2.5$ & $272\pm19$ \\
Fe~{\scriptsize II}~$\lambda4418$ & $25.6\pm8.8$ & $3.8\pm1.3$ & $295\pm112$ \\
He~{\scriptsize I}~$\lambda4472$ & $41.9\pm10.3$ & $6.0\pm1.5$ & $310\pm76$ \\
narrow H$\beta$ & $744.5\pm22.0$ & $91.5\pm2.7$ & $205\pm4$ \\
broad H$\beta$ & $210.1\pm68.3$ & $25.8\pm8.4$ & $2123\pm496$ \\
{[}O~{\scriptsize III}]~$\lambda4959$ & $1706\pm30$ & $208.9\pm3.7$ & $247\pm3$ \\
{[}O~{\scriptsize III}]~$\lambda5007$ & $5507\pm44$ & $673.1\pm5.4$ & $239\pm1$ \\
{[}Fe~{\scriptsize VII}]~$\lambda5159$ & $42.0\pm10.2$ & $5.5\pm1.3$ & $229\pm89$ \\
He~{\scriptsize I}~$\lambda5877$ & $139.7\pm11.9$ & $17.2\pm1.5$ & $218\pm14$ \\
{[}O~{\scriptsize I}]~$\lambda6302$ & $86.3\pm22.1$ & $9.4\pm2.4$ & $627\pm135$ \\
narrow H$\alpha$ & $2512\pm118$ & $221.0\pm10.4$ & $157\pm3$ \\
broad H$\alpha$ & $4795\pm165$ & $421.9\pm14.5$ & $1703\pm28$ \\
He~{\scriptsize I}~$\lambda6680$ & $39.3\pm10.1$ & $3.7\pm0.9$ & $192\pm46$ \\
{[}S~{\scriptsize II}]~$\lambda6717$ & $36.3\pm10.9$ & $3.5\pm1.1$ & $242\pm69$ \\
{[}S~{\scriptsize II}]~$\lambda6732$ & $22.6\pm14.9$ & $2.2\pm1.5$ & $170\pm94$ \\
He~{\scriptsize I}~$\lambda7065$ & $101.5\pm14.3$ & $10.7\pm1.5$ & $270\pm29$ \\
{[}Ar~{\scriptsize III}]~$\lambda7138$ & $41.1\pm14.0$ & $4.5\pm1.5$ & $252\pm64$ \\
{[}O~{\scriptsize II}]~$\lambda7322$ & $24.9\pm11.7$ & $3.0\pm1.4$ & - \\
{[}O~{\scriptsize II}]~$\lambda7332$ & $10.8\pm5.1$ & $1.3\pm0.6$ & - \\
O~{\scriptsize I}~$\lambda8446$ & $45.0\pm15.4$ & $6.3\pm2.1$ & $314\pm67$ \\
Pa10 & $35.2\pm16.1$ & $4.9\pm2.2$ & $222\pm77$ \\
{[}S~{\scriptsize III}]~$\lambda9701$ & $62.0\pm13.2$ & $8.6\pm1.8$ & $348\pm125$ \\
Pa9 & $53.5\pm12.3$ & $7.9\pm1.8$ & $206\pm110$ \\
{[}S~{\scriptsize III}]~$\lambda9533$ & $156.0\pm24.7$ & $24.5\pm3.9$ & $218\pm27$ \\
Pa$\epsilon$ & $69.4\pm15.2$ & $10.9\pm2.4$ & $199\pm129$ \\
Pa$\delta$ & $133.9\pm19.3$ & $22.6\pm3.3$ & $336\pm79$ \\
He~{\small I}~$\lambda10830$ & $530.0\pm37.5$ & $73.6\pm5.2$ & $245\pm18$ \\
\enddata
\tablecomments{Fluxes are not corrected for gravitational magnification. We show $3\sigma$ upper limits for non-detections.}
\label{tab:2476_opt_nir_line}
\end{deluxetable}

The strongest narrow line in the spectra of UNCOVER-2476 is [O~{\small III}]~$\lambda5007$. 
We measure an [O~{\small III}]~$\lambda5007$ flux of $5.51\pm0.04\times10^{-17}$~erg~s$^{-1}$~cm$^{-2}$, corresponding to an EW of $673\pm5$~\AA. 
Such a large [O~{\small III}]~$\lambda5007$ EW is among the upper $10\%$ of the values observed in the LRD population \citep{deGraaff2025c}. 
If the [O~{\small III}] emission primarily comes from the host galaxy, the large EW would indicate that the galaxy of UNCOVER-2476 is dominated by relatively young stellar populations. 

We may expect that the ionization-sensitive line ratios of UNCOVER-2476 are large given the strong [O~{\small III}] emission \citep[e.g.,][]{Tang2019,Sanders2020,Boyett2024}. 
We derive a very large [O~{\small III}]/[O~{\small II}] (O32) ratio ($37\pm3$). 
This value is more than $10\times$ of the typical O32 ratio of $z\simeq4$ star-forming galaxies ($\simeq3$; e.g., \citealt{Sanders2023,Shapley2023}),
and also $2\times$ larger than the average O32 measured from the composite spectrum of type I AGN at $z\simeq4-7$ ($19$; \citealt{Isobe2025}). 
We find a large [Ne~{\small III}]/[O~{\small II}] (Ne3O2) ratio as well, reaching Ne3O2 $=1.7\pm0.1$. 
These suggest that the narrow line emitting gas in UNCOVER-2476 is under extreme ionization conditions. 

The detection of auroral lines and [O~{\small III}]~$\lambda4959,5007$, [O~{\small II}]~$\lambda3727,3729$ emission enables us to jointly constrain the electron temperature and density of the narrow line emitting gas. 
The density-sensitive [O~{\small II}]~$\lambda3727,3729$ doublet is resolved in our G235M spectrum, with a doublet flux ratio of $f_{{\rm [OII]}\lambda3729}/f_{{\rm [OII]}\lambda3727}=1.1\pm0.2$. 
Together with the [O~{\small II}]~$\lambda\lambda7322,7332$ auroral lines and using the \texttt{PyNeb} code \citep{Luridiana2015}, we derive a temperature of $T_{\rm e}({\rm O}^+)=1.9^{+1.1}_{-0.7}\times10^4$~K and a density of $n_{\rm e}=470^{+310}_{-220}$~cm$^{-3}$ for the O$^+$ gas. 
For the O$^{2+}$ gas, we measure a [O~{\small III}]~$\lambda4363$/[O~{\small III}]~$\lambda5007$ ratio of $0.041\pm0.004$. 
If the O$^{2+}$ gas has the similar density as the O$^+$ gas, we derive an electron temperature of $T_{\rm e}({\rm O}^{2+})=2.2^{+0.1}_{-0.2}\times10^4$~K. 
Although note that the O$^{2+}$ gas density may not necessarily be the same as the O$^+$ gas, the temperature of O$^{2+}$ gas does not change significantly with density varying up to $\sim5\times10^4$~cm$^{-3}$. 

Based on the derived oxygen gas temperature and density, we can estimate the gas-phase oxygen abundance of UNCOVER-2476. 
Using the \texttt{PyNeb} code, we derive $12+\log{\rm (O/H)}=7.55^{+0.06}_{-0.05}$ ($Z=0.07^{+0.01}_{-0.01}\ Z_{\odot}$, where solar metallicity corresponds to $12+\log{\rm (O/H)}=8.71$; \citealt{Gutkin2016}). 
This is consistent with the average oxygen abundance of high redshift type I AGN inferred from [O~{\small III}]~$\lambda4363$ measurement ($12+\log{\rm (O/H)}=7.46$; \citealt{Isobe2025}). 
The result indicates that the narrow line emitting gas in UNCOVER-2476 is fairly metal poor.

\section{Rest-frame Optical Spectra of Abell2744-QSO1} \label{sec:QSO1_opt_spec}

\begin{figure*}
\includegraphics[width=\linewidth]{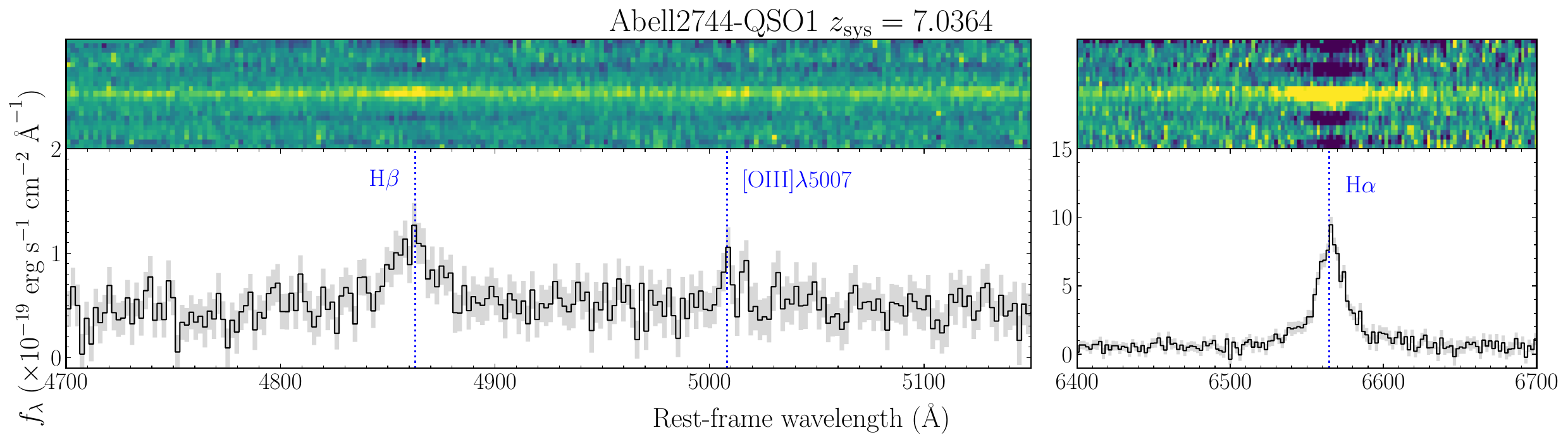}
\caption{SPURS JWST/NIRSpec rest-frame optical (G395M/F290LP and part of G235M/F170LP) spectra of Abell2744-QSO1, shown in the same way as Figure~\ref{fig:QSO1_uv_spec}. We detect broad H$\beta$ and H$\alpha$ emission lines, as well as a narrow, weak [O~{\scriptsize III}]~$\lambda5007$ emission line.}
\label{fig:QSO1_opt_spec}
\end{figure*}

We list the rest-frame optical emission line fluxes, EWs, and FWHMs measured from SPURS G395M spectrum of Abell2744-QSO1 in Table~\ref{tab:QSO1_opt_line}. 
The SPURS rest-frame optical spectrum of Abell2744-QSO1 is shown in Figure~\ref{fig:QSO1_opt_spec}.


\begin{deluxetable}{cccc}
\tablecaption{Rest-frame optical emission line flux ($\times10^{-20}$~erg~s$^{-1}$~cm$^{-2}$), EW (\AA), and FWHM (km~s$^{-1}$) of Abell2744-QSO1 measured from SPURS spectra.}
\tablehead{
Line & Flux & EW & FWHM
}
\startdata
{[}O~{\scriptsize II}]~$\lambda3728$ & $<35.6$ & $<20.9$ & - \\
{[}Ne~{\scriptsize III}]~$\lambda3869$ & $<19.0$ & $<7.6$ & - \\
H$\gamma$ & $<14.5$ & $<3.7$ & - \\
narrow H$\beta$ & $15.8\pm5.4$ & $3.2\pm1.1$ & $235\pm65$ \\
broad H$\beta$ & $152.9\pm32.5$ & $31.8\pm6.8$ & $1674\pm252$ \\
{[}O~{\scriptsize III}]~$\lambda4959$ & $<16.7$ & $<3.3$ & - \\
{[}O~{\scriptsize III}]~$\lambda5007$ & $25.4\pm7.9$ & $5.0\pm1.5$ & $251\pm80$ \\
narrow H$\alpha$ & $61.9\pm17.4$ & $11.2\pm3.1$ & $239\pm48$ \\
broad H$\alpha$ & $1069\pm264$ & $193.1\pm47.8$ & $2653\pm345$ \\
\enddata
\tablecomments{Fluxes are not corrected for gravitational magnification. We show $3\sigma$ upper limits for non-detections.}
\label{tab:QSO1_opt_line}
\end{deluxetable}


\bibliography{SPURS_LRD}{}
\bibliographystyle{aasjournalv7}



\end{document}